\documentclass[aps,prb,10pt,twocolumn,amsmath,amssymb,superscriptaddress,longbibliography,floatfix]{revtex4-2}
\usepackage{graphicx} 
\usepackage{bm}
\usepackage[usenames]{color}
\usepackage{amsmath}
\usepackage{placeins}

\newcommand{\be}{\begin{equation}}
\newcommand{\ee}{\end{equation}}
\newcommand{\beqn}{\begin{eqnarray}}
\newcommand{\eeqn}{\end{eqnarray}}

\begin{document}

\title{Quantum annealing of the random transverse-field Ising model in one, two, and three dimensions}

\author{Istv\'{a}n A. Kov\'{a}cs}
\email{istvan.kovacs@northwestern.edu}
\affiliation{Department of Physics and Astronomy, Northwestern University, 2145 Sheridan Road, Evanston, IL 60208.}
\affiliation{Department of Engineering Sciences and Applied Mathematics, Northwestern University, 2145 Sheridan Road, Evanston, IL 60208.}
\affiliation{Northwestern Institute on Complex Systems, Northwestern University, 600 Foster Street, Evanston, IL 60208.}

\author{Neil Shah}
\affiliation{Department of Physics and Astronomy, Northwestern University, 2145 Sheridan Road, Evanston, IL 60208.}

\author{R\'obert Juh\'asz}
\email{juhasz.robert@wigner.hun-ren.hu}
\affiliation{HUN-REN Wigner Research Centre for Physics, H-1525 Budapest, P.O.Box 49, Hungary}

\date{\today}

\begin{abstract}
Quantum annealing must be sufficiently slow to prevent defect formation. We study this problem in the random transverse-field Ising model via the adiabatic theorem to obtain the limiting rate at which defect formation can be avoided. Our approach based on the strong-disorder renormalization group (SDRG) technique views the annealing process as a gradual aggregation of strongly coupled clusters, giving access to the elementary events of error formation. 
We present a scaling theory of the statistics of these events in general dimensions and confront it to an extensive numerical analysis carried out in one, two, and three dimensions by an efficient implementation of the SDRG method. This approach is also validated by exact diagonalization in one dimension by means of the free-fermion technique. 
Our results consistently show a logarithmically slow decrease of the error density with the annealing time, characterized by the dimension-dependent critical exponents of the corresponding infinite-disorder fixed point of the model. Furthermore, we provide a map of individual events of potential defect formation in the excitation energy - control parameter plane, which may help in the design of optimal annealing paths. 
\end{abstract}

\maketitle

\section{Introduction}

Quantum annealing (QA) and adiabatic quantum computation aim to prepare the ground state of a target Hamiltonian by slowly evolving from a Hamiltonian with an easily prepared ground state~\cite{albash,hauke,kadowaki}. If the evolution is sufficiently slow, the adiabatic theorem~\cite{amin,albash} ensures that the system remains close to its instantaneous ground state. However, at finite annealing times, transitions to excited states can occur, particularly where the energy gap becomes small. These excitations produce errors in the final state and therefore impose a limit on the rate at which a system can be annealed successfully. Determining the minimum annealing time required to suppress such defects is consequently important for understanding both the performance of QA and the non-equilibrium dynamics of slowly driven quantum systems.

QA is often used to find the solutions of hard classical optimization problems~\cite{hauke, king}. This is achieved by encoding the cost function of the optimization problem as the ground-state energy of an Ising-type Hamiltonian, so that the ground state prepared at the end of the anneal directly yields the optimal (or near-optimal) solution~\cite{lucas}. Practically relevant optimization problems mapped in this way, such as max-cut and graph coloring~\cite{lucas}, tend to be heterogeneous: the coupling and local fields of the resulting Ising Hamiltonian are set by the specific problem instance and are therefore not translationally invariant. 

As a paradigmatic model of heterogeneous quantum systems, the random transverse-field Ising model (RTFIM) provides a useful setting for the purpose of understanding the fundamental limitations of QA protocols. Although the zero-field ferromagnetic ground state of the RTFIM is simple, deviations from this state can be identified as domain walls, providing a direct measure of defect production. In one dimension, the RTFIM can be mapped to free fermions, allowing its spectrum and relevant dynamical quantities to be calculated exactly~\cite{lsm,pfeuty}. The RTFIM therefore provides a setting in which approximate approaches to QA can be tested against exact calculations before being applied to systems where exact methods are unavailable.

Defect formation during a slow passage through a continuous phase transition is usually described by the Kibble-Zurek mechanism~\cite{kibble,zurek,cz,qKZ_zurek}, though known exceptions exist, such as the suppression of defect formation by a symmetry-breaking bias~\cite{rams2019, rysti2021} or by a spatially inhomogeneous quench~\cite{delcampo, dziarmaga_rams, gradient2013}. Near the critical point, the relaxation time of the system diverges and becomes larger than the time scale on which the Hamiltonian is changing. The state can then no longer follow the instantaneous ground state, resulting in a finite correlation length and corresponding density of defects. In the clean transverse-field Ising chain, this process can be studied exactly by decomposing the dynamics into independent Landau-Zener transitions~\cite{dziarmaga_prl,qKZ_zurek, qKZ_cherng}.

Quenched disorder substantially modifies this picture. The critical point of the RTFIM is governed by an infinite-disorder fixed point (IDFP) characterized by ultraslow, activated dynamical scaling~\cite{fisher}. At criticality, the characteristic energy scale decreases exponentially with a power of the length scale, causing the defect density to decrease only logarithmically with the annealing time~\cite{dziarmaga_random,caneva}. Disorder also produces broad distributions of excitation energies in the off-critical  Griffiths phases containing rare, arbitrarily low-energy excitations~\cite{fisher, im}. Nevertheless, the equilibrium spectrum alone does not identify which excitations control defect formation. The relevant annealing time scale is controlled by dynamically accessible excitations and their transition matrix elements appearing in the adiabatic theorem, rather than by the smallest equilibrium gap alone. A microscopic description of these excitations was recently developed using the strong-disorder renormalization group (SDRG)~\cite{mdh1980, fisher, im, im2018} in one dimension~\cite{juhasz2025}.

In the SDRG description, the annealing process is represented as a gradual aggregation of strongly coupled spin clusters~\cite{juhasz2025}. Each fusion of clusters acts as a local avoided crossing. If the passage is sufficiently slow, the merged cluster remains in its ground state, while a faster passage can leave it excited and produce a defect. Each fusion event can therefore be assigned a crossover annealing time separating the adiabatic and non-adiabatic regimes. For a given disorder realization, the largest of these crossover times sets the annealing time required to suppress defect formation.

The existing analytical treatment only applies to the infinite one-dimensional random transverse-field Ising chain~\cite{juhasz2025}. Applying this framework to finite systems and extending it to higher dimensions requires a numerical implementation of the renormalization-group procedure. Numerical SDRG studies have already determined the critical point, as well as the correlation-length exponent and tunneling exponent governing the transition, in two and three dimensions~\cite{kovacs2d,kovacs3d, Kovacs_2011}. These results provide part of the basis needed to formulate finite-size scaling predictions for the crossover annealing time in higher dimensions, though further ingredients are required.

The rest of the paper is organized as follows. In Sec.~\ref{sec:scaling}
 we introduce the model and the adiabatic-theorem framework used to define the crossover annealing time $\tau_c$, describe the SDRG-based microscopic theory of cluster fusion in one dimension, and generalize it to a scaling theory valid in higher dimensions. In Sec.~\ref{sec:results} we discuss our numerical techniques and present the numerical results, comparing first the free-fermion and SDRG results in one-dimension, and finding qualitatively similar SDRG results in two and three dimensions. We conclude with a discussion in Sec.~\ref{sec:discussion}. Details of the cluster-fusion and free-fermion derivations are presented in the Appendix.
 
\section{Quantum annealing and scaling expectations}
\label{sec:scaling}

\subsection{Formulation of the model}
\label{sub:model}

In this work, we consider the RTFIM with ferromagnetic nearest-neighbor couplings on $d$-dimensional cubic lattices, having the time-dependent Hamiltonian 
\be
{\cal H}(t) =
-\sum_{\langle ij\rangle}J_{ij}\sigma_i^x \sigma_j^x-\sum_ih_i(t)\sigma_i^z.
\label{Hgen}
\ee
Here, $\sigma_n^x$ and $\sigma_n^z$ denote Pauli operators,
the couplings $J_{ij}$ are assumed to be time-independent, while the 
transverse fields decrease monotonically to zero as time varies from $t=0$ to $t=\tau$, according to
\be 
h_i(t)=\left(1-\frac{t}{\tau}\right)g_0h_i^0.
\label{hit}
\ee 
The bare transverse fields $h_i^0$ and the couplings $J_{ij}$ are both assumed to be quenched, independent, identically distributed random variables with a uniform distribution in $[c_0,1+c_0]$, where $c_0\ge 0$. 
The parameter $g_0$ is chosen to be large, guaranteeing that the system is deeply in the paramagnetic phase at the beginning of the annealing ($t=0$). Then, during the annealing process the system is driven through its quantum critical point to the ferromagnetic phase. 
A frequently used control parameter of the model in Eq.~(\ref{Hgen}) is 
\be 
\theta(t)=\overline{\ln h_i(t)}-\overline{\ln J_{ij}},
\label{control}
\ee
where the overbar denotes an average over quenched disorder. With our parametrization, the control parameter varies with time as $\theta(t)=\ln(1-t/\tau)+\ln g_0$, with the critical control parameter given by $\theta_c$.

\begin{figure}[htbp!]   
    \centering
    \includegraphics[width=.95\linewidth]{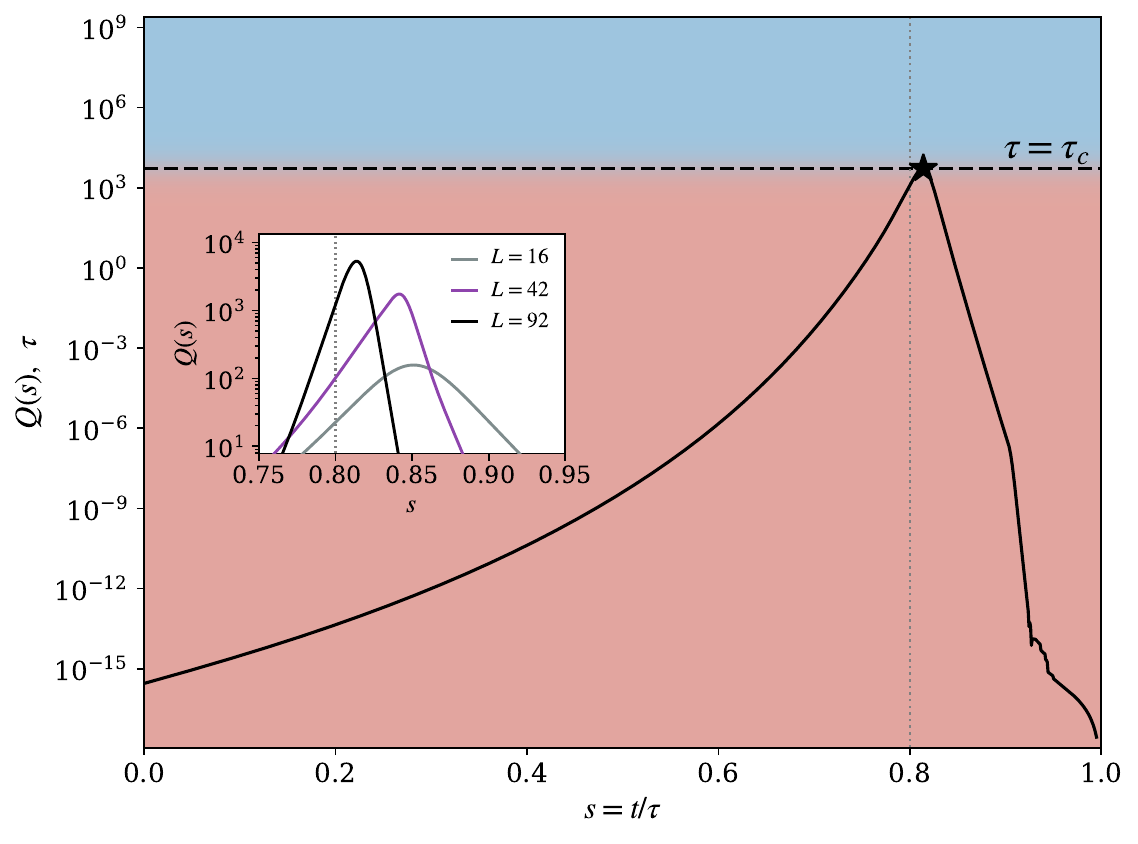}
\vskip-4mm
    \caption{\textbf{Illustration of the annealing-time threshold during QA.}  The data were obtained by the exact free-fermion calculation of $Q(s)$, 
    as described in Sec.~\ref{app:free-fermion}, for a single disorder realization of the $d=1$ RTFIM at size $L=92$. The anneal avoids  defects if $\tau \gg Q(s)$ for all $s$. The background shading, from red $(\tau \ll Q(s))$ to blue $(\tau \gg Q(s))$, reflects the crossover from a $\tau$ that would produce many defects to a $\tau$ that would produce few. The horizontal line marks $\tau_c = \max Q(s)$, which is the threshold annealing time of the sample. The vertical dotted line marks the critical point. Inset: $Q(s)$ near the peak for independent disorder realizations at $L = 16, 42$, and $96$.}
    \label{fig:Q_vs_s}
\end{figure}

In this work, rather than directly simulating the time evolution of the model, we aim at investigating the characteristics of error formation  during QA indirectly, through the adiabatic theorem~\cite{amin,albash}. 
According to the theorem, the annealing process is adiabatic, i.e.~the system will stay in its instantaneous ground state throughout the process, if the following condition for the annealing time $\tau$ is fulfilled: 
\be
\tau\gg\max_{0\le s\le 1}\frac{\left|\langle \psi_0(s)|\frac{d{\cal H}(s)}{ds}|\psi_1(s)\rangle\right|}{\epsilon^2(s)},
\label{at_gen}
\ee
Here, $s=t/\tau$ is the dimensionless time, $|\psi_0(s)\rangle$ and $|\psi_1(s)\rangle$ are the instantaneous ground state and first excited state, respectively, whereas $\epsilon(s)$ is the corresponding excitation energy. For any $t$, the Hamiltonian commutes with the parity operator $\hat P=\otimes_n\sigma_n^z$, which has eigenvalues $\pm 1$. Since the initial (ground) state is in the $+1$ parity sector, the time-evolved state must remain in this sector throughout the process. Therefore $|\psi_1(s)\rangle$ is meant to be the lowest-energy excited state in the $+1$ parity sector. 
Here, we focus on the threshold time scale $\tau_c$, defined as the r.h.s.~of the adiabatic theorem for a given Hamiltonian ${\cal H}(s)$ of finite size, 
\be
\tau_c\equiv \max_{0\le s\le 1} Q(s),
\label{tau_c}
\ee
where $Q(s)\equiv \left|\langle \psi_0(s)|\frac{d{\cal H}(s)}{ds}|\psi_1(s)\rangle\right|\epsilon^{-2}(s)$. Fig.~\ref{fig:Q_vs_s} illustrates this construction for a single disorder realization in one dimension, showing $Q(s)$ across the anneal together with the resulting $\tau_c$ and the critical point (vertical dotted line). The inset shows $Q(s)$ near its peak for samples of different sizes.

\subsection{Results in one dimension}
A microscopic description of the QA in the RTFIM based on the SDRG approach has been recently introduced and elaborated in one dimension~\cite{juhasz2025}. By the SDRG procedure performed at a fixed control parameter, the original spins of the model are classified into a set of clusters. This set completely determines the approximate ground state of the model: It is a product $|GS\rangle_{\rm SDRG}=\otimes_n|\uparrow_{C_n}\rangle$ of cluster states that are themselves ferromagnetic Greenberger-Horne-Zeilinger (GHZ) states of the constituting spins, $|\uparrow_{C_n}\rangle=\frac{1}{\sqrt{2}}\left(\otimes_{i\in C_n}|\rightarrow_i\rangle+\otimes_{i\in C_n}|\leftarrow_i\rangle\right)$. We note that the constituting spins of a given cluster do not necessarily form a spatially connected set.
When the control parameter is changed as in QA, from the paramagnetic toward the ferromagnetic phase, one can show in one dimension that clusters cannot split~\cite{juhasz2025}. Hence, the cluster set undergoes a gradual aggregation by pairwise fusions of clusters, which occur at certain values of the control parameter. 
Thus, the approximate SDRG ground state remains unchanged between subsequent fusion events. Therefore, during the annealing of the model, excitations arise predominantly at fusion events at which the SDRG ground state changes. 
For sufficiently slow (adiabatic) anneals, the product state of fusing clusters turns to a common ferromagnetic GHZ state, and the system remains in the ground state. Otherwise the fusion will be defective, the post-fusion state becoming an oscillating combination of a ferromagnetic GHZ state and an antiferromagnetic (domain-wall) state.

Within the SDRG approach, the fusion process of two clusters can be described by an effective two-spin transverse-field Ising model,
\be
\mathcal{H}_{\rm 12}(t)=-\tilde J(t)\tilde\sigma_1^x\tilde\sigma_2^x-\tilde h_1(t)\tilde\sigma_1^z -\tilde h_2(t)\tilde\sigma_2^z.
\ee
The time $t^*$, marking the fusion point, is determined by the condition $\tilde J(t^*)=\tilde h_1(t^*)$.
Here and in the following, without loss of generality, we assume that $\tilde h_1(t^*)>\tilde h_2(t^*)$.
It was pointed out earlier~\cite{juhasz2025} that $t^*$ essentially coincides with the point at which the ratio $Q(s)$ in the adiabaticity condition is maximal. 
As it is shown in Appendix~\ref{maximum}, the crossover annealing time can be written as  
\be 
\tau_c \sim \frac{n_1+n_b}{\epsilon(t^*)},
\label{taueps}
\ee
where $n_1$ and $n_b$ denote the number of original fields involved in $\tilde h_1$ and $\tilde J$, respectively. 
For brevity, we will call $n_1+n_b$ the multiplicity of the fusion.
The considerations quoted up to this point are generally valid for dimensions $d\ge 1$. 

Specifically, in the one-dimensional model, the analytic treatment leads to the following results~\cite{juhasz2025}. 
The conditional probability distribution of the logarithmic energy scale $\Gamma\equiv -\ln\epsilon$ of fusions occurring at a given value of the control parameter $|\Delta| = |\theta - \theta_c|\ll 1$ has an exponential cutoff at small energy (large $\Gamma$) as $G_>(\Gamma|\Delta)\sim |\Delta|e^{-2|\Delta|\Gamma}$.
This means that an $O(1)$ constant fraction of fusions falls in the range $\Gamma<\Gamma_{\Delta}$, where the $\Delta$-dependent threshold $\Gamma_{\Delta}$ diverges at the critical point as 
\be 
\Gamma_{\Delta}\sim |\Delta|^{-1}.
\label{gammadelta}
\ee
Integrating over the entire control parameter range of the annealing process yields that the fraction of erroneous (non-adiabatic) fusions scales, for large annealing times, as 
\be 
n(\tau)\sim \left[\ln\frac{\tau/\tau_0}{\ln^2(\tau/\tau_0)}\right]^{-2}.
\ee

\subsection{Generalization to higher dimensions}
\label{sub:gen}

In dimensions $d>1$, the picture of viewing the annealing process as a series of cluster fusions is contingent on the validity of the SDRG method, i.e.~on the RTFIM flowing to an IDFP in the dimension considered. SDRG studies have confirmed this to be the case up to $d = 4$~\cite{kovacs3d, Kovacs_2011}. Although no numerical results are available above $d = 4$, it has been conjectured that discrete-symmetry systems such as the RTFIM have an IDFP in any finite dimension provided the initial disorder is strong enough~\cite{im}.
Under this assumption, the relation in Eq.~(\ref{taueps}) obtained for the two-cluster fusion process is generally valid for any dimension $d \ge 1$, apart from factors that are $O(1)$.
Now, we consider a finite system with linear size $L$ and the maximal $\tau_c$ among all fusion events of the sample. This quantity, $\tau_L$, which approximates $\tau_c$ of Eq.~(\ref{tau_c}) evaluated for the full Hamiltonian of the sample, determines the minimal annealing time required for the process to be adiabatic, i.e.~to end up with the ground state of the system with a high probability. 

\begin{table}[htbp!]
\centering
\caption{
\textbf{Critical parameters of the RTFIM.} The number in the parentheses gives an error estimate of the last digits. Note that for $h$ we use the interval $(0.1,1.1)$ in $d=1$, and $(0,1)$ in $d=2$ and  $d=3$. The last two lines list the conjectures presented in the main text.
 \label{table:1}}
\begin{tabular}{cccc}  
 $d$  & 1 \cite{fisher} & 2 \cite{kovacs2d}& 3 \cite{kovacs3d, Kovacs_2011}\\ 
   \hline\hline
    $\theta_c$  & $0$  & $ 1.6784(1)$ &$2.5305(10)$ \\
$\psi$    & $0.5$  & $ 0.48(2)$ &$0.46(2)$ \\ 
$\nu$    & $2$ & $1.24(2)$ &$0.987(17)$  \\
$d_f$    & $\frac{\sqrt{5}+1}{4}\approx0.809$ & $1.018(15)$  &$1.160(15)$ \\
$\omega\,\,\,\, (> d_f)$    & $1$ & $\approx1.28$  &$\approx 1.4$ \\
$\omega=\psi+1/\nu$    & $1$ & $1.29(4)$  &$1.47(4)$ \\
$\kappa= 1+d/\psi$  & $3$ & $5.2(2)$  &$7.5(3)$ 
  \end{tabular}
  \end{table}

To obtain the scaling of $\tau_L$ with the system size $L$, we can combine the finite-size scaling of the gap and of the multiplicity at the IDFP governing the transition. 
Here, the effective couplings and fields of the renormalized system are known to follow an activated scaling with the length scale~\cite{fisher, Motrunich2000, kovacs2d, kovacs3d}. As a consequence, the minimal energy gap $\epsilon(t^*)$ in the even parity sector, denoted from now on simply by $\epsilon$ also obeys an activated dynamical scaling of the form~\cite{fisher,tang},
\begin{equation}
    \epsilon \sim e^{-cL^\psi}
    \label{eps_L}
\end{equation}
with the tunneling exponent $\psi$ given for $d = 1,2,3$ in Table~\ref{table:1}. 
Here, $c$ denotes an $O(1)$ random variable, so that the reduced variable $x_{\epsilon}=\Gamma L^{-\psi}$ has a non-trivial limit distribution as $L\to\infty$. 
Concerning the multiplicity $n_1+n_b$ of the dominant fusion event, we note that a slightly different quantity called the size of energy cluster has been introduced and studied in an earlier work~\cite{sdrg_k12}. As this related quantity was found to scale algebraically with the system size at the IDFP in different dimensions, we anticipate an algebraic growth of the multiplicity of the dominant fusion event with the system size,
\begin{equation}
    n_1+n_b \sim L^\omega,
    \label{n_L}
\end{equation}
characterized by the same exponent $\omega$, the numerical estimates of which taken from Ref.~\cite{sdrg_k12} are listed in Table~\ref{table:1}.

We conjecture that $\omega$ is not independent from the usual critical exponents, but can be expressed as a combination of $\psi$ and $\nu$. This statement relies on the following heuristic argument. 
For a finite sample of linear size $L$, let us consider the control parameter of the sample obtained by evaluating the averages in Eq.~(\ref{control}) for couplings and fields of the sample. 
Specially, in $d=1$ with periodic boundary condition, where we have as many fields as couplings at any stage of the SDRG procedure, the control parameter can be written as 
\be 
\Delta_L=\frac{\sum_i\ln\tilde h_i-\sum_i\ln\tilde J_i}{\sum_i n_i+ \sum_i n_{b,i}}.
\label{Delta_L}
\ee
Here, $n_i$ ($n_{b,i}$) denote the number of original fields involved in the $i$th effective field (bond), with $n_{b,i}\equiv n_{i,i+1}$ in the notation of Sec.~\ref{subsection:numerical_techniques} (cyclic indices).
Initially, $n_i=1$ and $n_{b,i}=0$ for $i=1,\dots,L$. Indeed, both the numerator and the denominator in Eq.~(\ref{Delta_L}) remain invariant under renormalization, the latter being $L$. We note, that in $d=1$, $\theta_c=0$ or, equivalently $\Delta_L\to 0$ as $L\to\infty$.
For $d>1$, where there are more couplings than fields, such an invariant combination is not known. Here, the interaction network of effective spins (clusters) will quickly develop under renormalization from the initial regular lattice structure to a complete graph~\cite{Motrunich2000,kovacs2d}. This will, however, return to a cyclic arrangement similar to the case $d=1$ close to the end of the renormalization procedure, when there are only three clusters left. 

At this stage, we can apply Eq.~(\ref{Delta_L}) to define a proper (reduced) control parameter of the sample, which tends to zero as $L\to\infty$ for the critical ensemble with $\theta=\theta_c$.  We then perform one more decimation step, which leaves just two clusters with fields $\tilde h_1$ and $\tilde h_2$ in a ``cyclic" arrangement, i.e.~with two couplings $\tilde J_1$ and $\tilde J_2$. If these couplings are treated formally as distinct, the control parameter $\Delta_L$ does not change. At this stage, we make use that at an IDFP the renormalized (logarithmic)  couplings and fields have broad distributions for large $L$, and $\tilde h_1\gg \tilde h_2$ and $\tilde J_1\gg \tilde J_2$ is fulfilled with a high probability with increasing $L$. 

Now, let us consider the sample at time $t=t^*$ when the dominant fusion occurs (between the last two remaining clusters). Here, we have thus $\tilde J_1\approx\tilde h_1$, and the control parameter can be written as 
\be 
\Delta_L\approx \frac{\ln\tilde h_2-\ln\tilde J_2}{n_1+n_2+n_{b,1}+n_{b,2}}.
\label{Delta2}
\ee
As the energy gap within the even sector is approximately minimal here, this point can be regarded as a pseudocritical point of the (finite) sample. As such, we expect its magnitude to scale with $L$ in the same way as control parameters at finite-size pseudocritical points do, i.e.~as $|\Delta_L|\sim L^{-1/\nu}$. 

The relevant 
energy gap of the sample is approximately given by twice the effective field $\tilde h_2$, so the first term in the numerator of Eq.~(\ref{Delta2}) scales with $L$ as $|\ln\tilde h_2|~\sim L^{\psi}$. As the renormalization procedure generally drives the model in any dimension to a closed chain of three effective spins at the final stage, we expect an asymptotic duality between the effective couplings and fields at criticality. Consequently, we expect the second term in the numerator, $-\ln\tilde J_2$, to scale with $L$ in the same way as the first one.
Furthermore, the denominator in Eq.~(\ref{Delta2}) is the sum of the multiplicity of the dominant fusion $n_1+n_{b,1}\equiv n_1+n_b$ and of a virtual fusion $n_2+n_{b,2}$. Thus, it scales with $L$ by definition as $\sim L^{\omega}$. 
The consistency between the above finite-size scaling forms of the factors in Eq.~(\ref{Delta2}) requires then the following exponent relation to hold:
\be 
\omega=\psi+\frac{1}{\nu}.
\label{conj}
\ee
This relation is satisfied by the exact critical exponents of the $d=1$ system, and is also compatible with the numerical estimates in $d=2$ and $3$ given in Table~\ref{table:1}. 
Another argument supporting Eq.~(\ref{conj}), based on the distribution of excitation energies of fusions, is presented at the end of this section.

To obtain the finite-size scaling of the threshold annealing time $\tau_L$, we substitute Eqs.~(\ref{eps_L}) and~(\ref{n_L}) into Eq.~(\ref{taueps}), yielding
\begin{equation}
    \tau_L \sim e^{cL^{\psi}}L^{\omega},
    \label{tau_L}
\end{equation}
or, in terms of logarithmic variables, $\ln{\tau_L} \sim cL^\psi + \omega \ln{L}$. This implies that the reduced variable $\ln(\tau_L)L^{-\psi}$ has the same limit distribution as $x_{\epsilon}$. Nevertheless, the convergence to this limit distribution is somewhat slower for $\ln(\tau_L)L^{-\psi}$, as deviations at finite sizes are of $O(L^{-\psi}{\ln L})$ rather than $O(L^{-\psi})$ valid for $x_{\epsilon}$. 

Besides the minimal annealing time required to perform QA free from errors in a sample of finite size, another (related) characteristic of the effectiveness of QA is the density of errors produced in an infinite system under a finite annealing time. 
We note that a single erroneous fusion of two clusters appears in the final state as a domain wall between two ferromagnetically ordered domains the cores of which are the two clusters. Apart from the one-dimensional system for which the domain-wall is point-like, the shape of domain walls in higher dimensions is non-trivial. 
For the sake of simplicity, we will not deal here with the question how the final state is shaped by erroneous fusions, and measure the density of errors by the fraction $n(\tau)$ of erroneous (non-adiabatic) fusion events to all fusion events. A practical criterion for classifying a fusion event to be erroneous is that its characteristic crossover time $\tau_c$ exceeds the annealing time $\tau$. 

To obtain the scaling of $n(\tau)$ with $\tau$, we invert the relationship in Eq.~(\ref{tau_L}), resulting in 
$L^{\psi} \sim \ln{\tau} - \frac{\omega}{\psi}\ln{\ln{\tau}}$, where we kept the first sub-leading term near the leading term $\ln \tau$ for large $\tau$.
This inverted relationship gives the scaling of the typical linear size of locally error-free (ferromagnetic) domains in an infinite system with annealing time $\tau$. 
Thus, the scaling of error density can be obtained by substituting this relationship to 
$n(\tau)\sim L^{-d}$, yielding 
\beqn
n(\tau) \sim \left(\ln{\frac{\tau}{(\ln{\tau})^{\frac{\omega}{\psi}}}}\right)^{-\frac{d}{\psi}},
\eeqn
or, equivalently, 
\beqn
[n(\tau)]^{-\psi/d} \sim \ln{\tau} - \frac{\omega}{\psi}\ln{\ln{\tau}}.
\label{eq:tau_c_n_scaling}
\eeqn

Next, we consider the envelope $\Gamma_{\Delta}$ bordering the domain in the $1/\Gamma-\Delta$ plane, which contains an $O(1)$ fraction of fusion events. The shape of this curve at $\Delta=0$ in general dimensions, which is given for $d=1$ in Eq.~(\ref{gammadelta}), can be obtained as follows. 
Outside of the critical point, the average spatial correlation length $\xi$ is finite and diverges close to the critical point as $\xi\sim |\Delta|^{-\nu}$, with the (average) correlation-length exponent $\nu$ listed in Table~\ref{table:1}. The length $\xi$ can be interpreted in the SDRG procedure as a characteristic renormalization length scale up to which the system is essentially critical, meaning that field and bond decimations occur on a comparable rate. Beyond this length scale either of the decimation types will dominate the other, and, consequently, fusion events become rare. With $\xi$, a (logarithmic) energy scale $\Gamma_{\Delta}$ can be associated through the relation $\Gamma_{\Delta}\sim \xi^{\psi}$ valid at the IDFP, since the system is critical within this scale. Thus, we obtain that the domain containing an $O(1)$ fraction of fusion events is limited by an envelope of the shape at $\Delta=0$
\be 
\Gamma_{\Delta}\sim |\Delta|^{-\nu\psi}.
\label{eq:envelope}
\ee

In the final part of this section, we discuss the distribution of excitation energies of all fusions occurring during the entire annealing process. 
We start with the joint probability density $g(\Gamma,\Delta)$ of logarithmic excitation energy $\Gamma$ and reduced control parameter $\Delta$ of fusions. In $d=1$, it was established in Ref.~\cite{juhasz2025} that the probability density (close to the critical point) is composed of the following factors 
\beqn
g(\Gamma,\Delta)\, d\Gamma\, d\Delta \sim \Theta_0(\Gamma,\Delta)\,\pi_0(\Gamma,\Delta) \nonumber \\
\times\, m(\Gamma,\Delta)\, n_\Gamma(\Delta)\, d\Gamma\, d\Delta.
\label{joint}
\eeqn
Here, $\Theta_0$ and $\pi_0$, originating in the SDRG treatment of the model, are the functional values of the distributions of logarithmic variables, $\beta=\ln(\Omega/h)$ and $\zeta=\ln(\Omega/J)$ at the upper edge $h=\Omega$ and $J=\Omega$, respectively. $\Omega\equiv e^{-\Gamma}$ denotes the running SDRG energy scale, the largest remaining field or coupling at a given stage of renormalization, decreasing monotonically from its initial value $\Omega_0$ as decimation proceeds. $m(\Gamma,\Delta)$ denotes the mean multiplicity of fusions at renormalization scale $\Gamma$ and control parameter $\Delta$. The rationale for the appearance of $\Theta_0\pi_0m$ in Eq.~(\ref{joint}) is that a fusion corresponds to a crossing of a field and a neighboring coupling by an infinitesimal change of $\Delta$. The factor $\Theta_0$ accounts for the weight of finding a field to be decimated, while $\pi_0m$ is the weight of finding a neighboring coupling with a $\zeta$ being in a zone of width $\sim md\Delta$ at $\zeta=0$.
Here, the factor $m$ is included because the shift $d\Delta$ results in a different amount of individual shifts of renormalized fields and bonds, depending on their multiplicities. In other words, $m$ gives the mean speed of crossings as $\Delta$ is varied, which depends on $\Gamma$ and $\Delta$. 
Finally, the factor $n_\Gamma(\Delta)$ is the ratio of the number of active (non-decimated) clusters at scale $\Gamma$ to the number of original spins, taking into account that the density of effective spins decreases as the renormalization progresses.

The decomposition of $g(\Gamma,\Delta)$ in Eq.~(\ref{joint}) is reasonable in higher dimensions, as well, although we have to properly take into account how each of the factors is affected by the dimensionality of the system. 
We note that in $d=1$, the marginal probability density of $\Delta$ of fusion events is constant at the critical point $\Delta=0$ in leading order, which is implicit in Eq.~(\ref{joint}), and we assume this to be valid also in higher dimensions. 

Now, we are interested in the probability density $\rho(\Gamma)$ of the logarithmic excitation energies of fusion events occurring at any stage of the annealing procedure. This is nothing but the marginal probability density $\rho(\Gamma)=\int g(\Gamma,\Delta)d\Delta$. 
We restrict ourselves to the large-$\Gamma$ tail of $\rho(\Gamma)$, which we anticipate to be algebraic, 
\be 
\rho(\Gamma)\sim \Gamma^{-\kappa},
\label{tail}
\ee
and focus on the exponent $\kappa$. Two simplifications are then possible. i) It is sufficient to restrict the integral over $\Delta$ to the range $|\Delta|<\Delta^*\sim\Gamma^{-1/(\nu\psi)}$ since, out of this range (the envelope $\Gamma_{\Delta}$), fusions are negligibly rare. ii) Within this range, the renormalization flow is critical, so that we can use the forms of all constituting factors in Eq.~(\ref{joint}) valid at the critical point $\Delta=0$. 
Thus, within the above range, the critical relationship between energy and length scale 
\be 
\Gamma\sim \xi^{\psi}
\label{Gamma_xi}
\ee
holds, which, using $n_\Gamma\sim \xi^{-d}$, leads to $n_\Gamma\sim \Gamma^{-d/\psi}$.
Furthermore, we have $\Theta_0\sim \pi_0\sim 1/\Gamma$ in general dimensions, which is necessary to obtain an algebraic relationship in Eq.~(\ref{Gamma_xi}), the dimension affecting only the prefactors. 
Finally, we have $m(\Gamma)\sim \Gamma^{\omega/\psi}$ by Eqs.~(\ref{n_L}) and (\ref{Gamma_xi}). 
Substituting these forms into Eq.~(\ref{joint}) and performing the integral from $-\Delta^*$ to $\Delta^*$, results in an algebraic tail of $\rho(\Gamma)$ as given in Eq.~(\ref{tail}), with the exponent
\be
\kappa=2+(d-\omega+1/\nu)/\psi. 
\label{kappa1}
\ee
Let us consider first the case $d=1$, where we have $\kappa=3$. Here, the dominant fusion of a system of size $L$ is the fusion having approximately the largest $\Gamma$ (smallest excitation energy) among the $\Gamma$ variables of all fusions ($L$ in number) of the sample.
Although the $\Gamma$ variables of different fusions of a given sample are not strictly independent, the known finite-size scaling of the maximum, $\Gamma_L\sim L^{1/2}$ is in agreement with the prediction of extreme-value statistics valid for independent variables: $\Gamma_L\sim L^{1/(\kappa-1)}$.   
Under the assumption that the finite-size scaling law of independent variables holds to be asymptotically valid also in higher dimensions, the scaling $\Gamma_L\sim L^{\psi}$ of the maximum out of $L^d$ variables requires the tail exponent to be 
\be
\kappa=1+d/\psi.
\label{kappa2}
\ee
Equating this expression to that in Eq.~(\ref{kappa1}), leads us again to the conjecture in Eq.~(\ref{conj}). 

Using Eq.~(\ref{tau_L}), we conclude that the probability density of logarithmic crossover times $\ln\tau_c$ of all fusions must have, apart from corrections, the same algebraic tail as that of $\Gamma$ has. 
By similar arguments relying on extreme-value statistics, we expect the probability density of the multiplicities $n_1+n_b$ of all fusions to have an algebraic tail 
\be
\rho(n_1+n_b)\sim (n_1+n_b)^{-1-d/\omega}
\label{mtail}
\ee
for large multiplicities.

\section{Numerical results}
\label{sec:results}

\subsection{Numerical techniques}
\label{subsection:numerical_techniques}

To start, for the one-dimensional model, we applied the free-fermion technique described in Appendix~\ref{app:free-fermion}, which maps the Hamiltonian in Eq.~(\ref{H}) with open boundary condition to a quadratic form in Jordan-Wigner fermions. This reduces the diagonalization to a singular-value decomposition of the bidiagonal matrix $M(s)$ in Eq.~(\ref{M}). The resulting single-particle excitation energies $\epsilon_i(s)$ give the exact instantaneous many-body gap within the even sector, $\epsilon(s)=\epsilon_1(s)+\epsilon_2(s)$, while the matrix element entering the numerator of Eq.~(\ref{at_gen}) is calculated from the components of singular vectors. This reduces the problem to operations on an $L \times L$ matrix rather than the full $2^L$-dimensional Hilbert space, reaching system sizes well beyond direct diagonalization. 
Using the parameters $g_0=5$ in Eq.~(\ref{hit}) and $c_0=0.1$, we scanned $s=t/\tau$ from $0$ to $1$ with an initial resolution $0.01$, evaluated $Q(s)$ at each point, and determined the maximum $\tau_L=Q(s^*)$ by successive approximation, as illustrated in Fig.~\ref{fig:Q_vs_s}. We also evaluated the energy gap $\epsilon(s^*)$ at this point. We repeated this protocol for system sizes up to $L=256$, with over $10^5$ independent realizations of disorder for each size. From these data, we calculated the distributions of $\tau_L$, $\epsilon$, and $\tau_L\epsilon$; the latter separates out the matrix element in the numerator of $Q(s^*)$. 

\begin{table}[htbp!]
\centering
\caption{
\textbf{Critical window in various dimensions.} The window size is estimated from the known $\theta_c$ and $\nu$ values from Table~\ref{table:1} using $c_{\pm}L^{-1/\nu}$, with a sufficiently large $c_{\pm}$ prefactors on each side so that all data collected falls in this window. 
 \label{table:window}}
\begin{tabular*}{\columnwidth}{@{\extracolsep{\fill}}lcc|lcc|lcc}
    &$d=1$ &   &  & $d=2$ & &  & $d=3$  & \\ 
    \hline
  $c_{\pm}$   & $10$ &  $10$ &   & $13$ & $19$ &   & $17$ & $10$\\ \hline\hline
     $l$ & $\theta_{\mathrm{min}}$& $\theta_{\mathrm{max}}$& $l$ & $\theta_{\mathrm{min}}$& $\theta_{\mathrm{max}}$& $l$  & $\theta_{\mathrm{min}}$& $\theta_{\mathrm{max}}$\\ 
   \hline
512&-0.442&0.442&24&0.676&3.143&8&0.463&3.747\\
1024&-0.313&0.313&32&0.884&2.840&12&1.159&3.337\\
2048&-0.221&0.221&48&1.105&2.516&16&1.506&3.133\\
4096&-0.156&0.156&64&1.224&2.342&24&1.851&2.930\\
8192&-0.110&0.110&96&1.351&2.157&32&2.023&2.829\\
16384&-0.078&0.078&128&1.419&2.058&48&2.194&2.728\\
32768&-0.055&0.055&192&1.491&1.952&64&2.279&2.678\\
65536&-0.039&0.039&256&1.530&1.895&96&2.364&2.629\\
131072&-0.028&0.028&384&1.571&1.835&128&2.406&2.604
  \end{tabular*}
  \end{table}

As a more scalable alternative, which also works in higher dimensions, we generalized the SDRG technique to evaluate $Q(s^*)$.
The SDRG procedure successively eliminates the high-energy excitations following a perturbative decimation scheme for random quantum spin systems~\cite{mdh1980, im}. Here, two kinds of reduction steps are applied, each of which gradually decreases the number of effective spins and the energy scale $\Omega$ set by the actually largest local parameter (bond or a field). When the largest parameter is a transverse field, $\Omega=\tilde h_i$, the corresponding cluster is integrated out, generating an effective coupling $\tilde J_{jk} = \tilde J_{ij}\tilde J_{ik}/\tilde h_i$ between neighboring clusters via second-order perturbation theory. Conversely, decimating a coupling, $\Omega=\tilde J_{ij}$, merges the two strongly interacting clusters into a single ferromagnetic unit. Its low-energy states are described by an effective spin-$1/2$ degree of freedom subject to a transverse field $\tilde h_{ij}=\tilde h_i\tilde h_j/\tilde J_{ij}$~\cite{fisher}. In two and higher dimensions, we use the maximum rule, justified by the infinitely broad parameter distributions at the critical point, retaining only the strongest coupling when multiple bonds connect the same clusters~\cite{kovacs2d,kovacs3d,Kovacs_2011}. Iterating these decimation steps in a finite system until all clusters are eliminated, we obtain the cluster configuration and the corresponding effective parameters at a given value of the control parameter.

To investigate QA, screening through all control parameter values would in principle require a time-consuming process of updating the control parameter in sufficiently small steps such that no cluster-fusion event is missed. This can be, however, circumvented by estimating at each control parameter value the precise value at which the next event would occur.

For practical reasons, explained later, we explored the variation of cluster set in a reversed order, i.e.~moving from the ferromagnetic phase to the paramagnetic one.  
Therefore, we initialize the SDRG procedure with a sufficiently small external field such that all sites form a single cluster. 

The effective transverse field of a cluster appears in the generic form of $\tilde h_i=\prod_{x\in\mathcal{F}_i} h_x/\prod_{y\in\mathcal{B}_i}J_y$. 
Here, $\mathcal{F}_i$ and $\mathcal{B}_i$ denote the multisets of (not necessarily distinct) original fields and couplings merged into $\tilde h_i$,
while $n_i \equiv |\mathcal{F}_i|$ is the number of such fields, counted with multiplicity. Similarly, each effective coupling is given in the form $\tilde J_{ij}=\prod_{u\in \mathcal{B}_{ij}}J_u/\prod_{v\in \mathcal{F}_{ij}}h_v$, with $n_{ij} \equiv |\mathcal{F}_{ij}|$.
Therefore, upon changing the $\theta\equiv \log \lambda$ control parameter, such that each original field is transformed as $ h'=\lambda h$, we change the effective parameters as $\tilde h'_i= \tilde h_i \lambda^{n_i}$ and $\tilde J'_{ij}= \tilde J_{ij} \lambda^{-n_{ij}}$.
From the condition of balancing out the next potential site decimation or merging events for each site $i$, $\tilde h'_i=\tilde J'_{ij}$, it follows that the next $\theta$ control parameter value of interest can be estimated as  
\be
\theta^*=\max_{i, j} \frac{\log \tilde J_{ij}-\log \tilde h_i}{n_i+n_{ij}},
\label{thetastar}
\ee
where the index $j$ runs through the current neighbors of site $i$. For the pair of clusters $i$ and $j$ fusing through $\tilde J_{ij}$, with $\tilde h_i$ being the larger field, the multiplicities are the same as in Sec.~\ref{sec:scaling} denoted by $n_1 \equiv n_i$ and $n_b \equiv n_{ij}$.

Note that the formula in Eq.~(\ref{thetastar}) can underestimate the control parameter change at which the next event occurs. Such events are extremely rare in $d>1$, but in $d=1$ appear to happen in a finite fraction of the cases, leading to a constant multiplier on the number of iteration steps needed. Note that this SDRG protocol can never overestimate the location of the next relevant control parameter value, indicating that no relevant excitation can be missed in the process. In the end, the excess estimates where the number of clusters does not increase are left out of the analyses, as well as the invalid events where $n_b=0$ and $n_1=1$, see Appendix~\ref{maximum}. 
We also note that the above mentioned SDRG protocol could not work if the field was decreased from a large initial value, as the current SDRG architecture does not reveal what the next merging event would be. Nevertheless, once the cluster-splitting events are uncovered, we can equivalently consider the process in terms of a decreasing external field.

As long as only the dominant excitation, i.e.~the one having the largest crossover time, $\tau_L$, is concerned, the SDRG calculations can be further reduced by limiting the search to the critical region. The benefit of such a focused search increases quickly with the system size, as the size of the critical window is expected to shrink asymptotically as $c_{\pm} L^{-1/\nu}$. We chose the prefactors $c_{-}$ and $c_{+}$ on the two sides of the critical point conservatively, such that all threshold data collected with the full search falls well within the corresponding window as listed in Table~\ref{table:window}.

We performed the numerical SDRG calculations in dimensions $d=1,2$, and $3$, using periodic boundary condition. The offset in the initial distribution of bare fields was $c_0=0.1$ in $d=1$ and $c_0=0$ otherwise. As described in Appendix~\ref{maximum}, we determined the approximate point of maximum $s^*$ under the approximation that the field of one of the fusing clusters is much smaller than the other. 
These calculations were performed typically for $10^5$ but at least a few times $10^4$ samples for each system size. 
Then, as by the exact-diagonalization technique, we determined the distributions of $\tau_L$, $\epsilon$, and $\tau_L\epsilon$. We also computed the distribution of the multiplicity $n_1+n_b$ to estimate the exponent $\omega$ directly. 
Furthermore, we considered the distributions of the above quantities associated with all fusions (not only the dominant one) of a large sample (or more), and the joint distribution of the excitation energy and control parameter of fusions.

\subsection{Validation of the SDRG method in one dimension}

\begin{figure*}[htbp!]   
    \centering
    \includegraphics[width=0.32\linewidth]{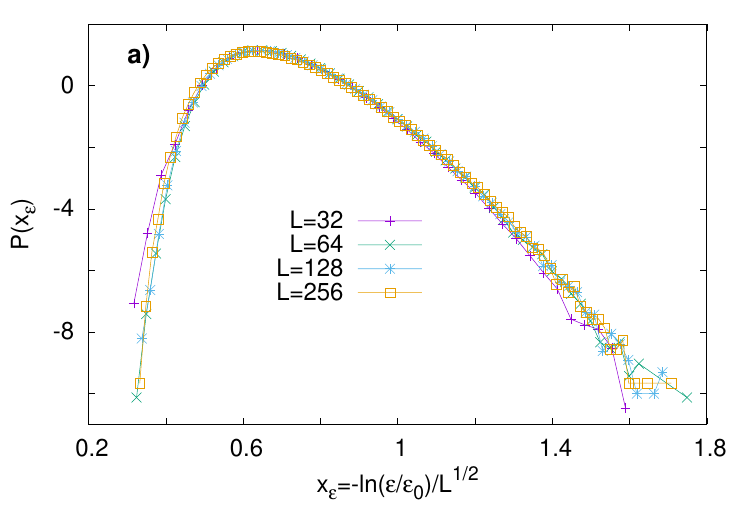}
    \includegraphics[width=0.32\linewidth]{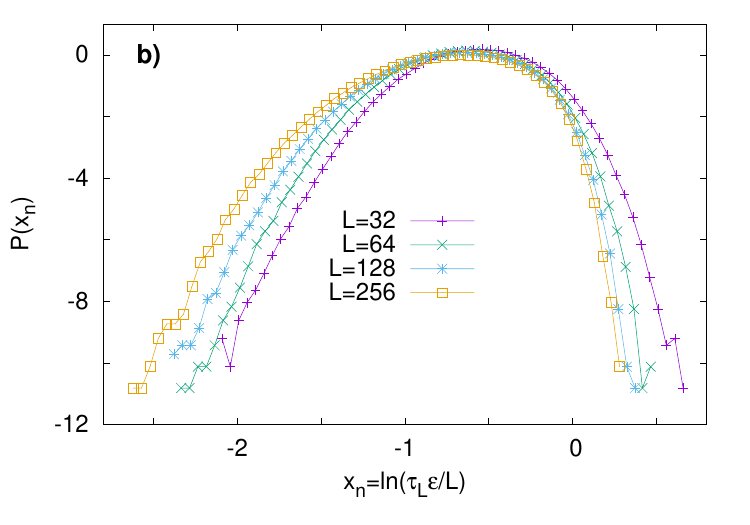}
    \includegraphics[width=0.32\linewidth]{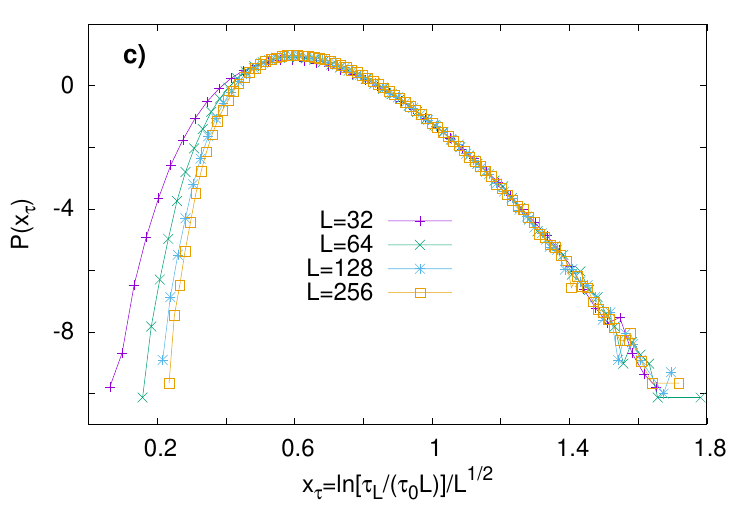}
    \vskip-5mm
     \caption{\textbf{Scaling collapse of distributions of exact-diagonalization data in one dimension.} a) excitation energy, $x_{\epsilon} = -\ln(\epsilon/\epsilon_0)/L^{1/2}$ with parameter $\epsilon_0=2$; b) effective multiplicity, $x_n=\ln(\tau_L\epsilon/L)$; c) threshold annealing time, $x_{\tau}=\ln[\tau_L/(\tau_0L)]/L^{1/2}$ with parameter $\tau_0=0.4$.}
     \label{fig_ed}
\end{figure*}

As detailed in Appendix~\ref{app:free-fermion}, we used exact diagonalization to produce reference distributions of the threshold annealing time $\tau_L$, the excitation energy $\epsilon$ at $s^*$, and the combination $\tau_L\epsilon$. According to the scaling behavior summarized in Eqs.~(\ref{eps_L}), (\ref{n_L}) and~(\ref{tau_L}), the distributions of scaling combinations $x_{\epsilon}=-\ln(\epsilon/\epsilon_0)/L^{\psi}$, $x_n=\ln(\tau_L\epsilon/L^{\omega})$, and $x_{\tau}=\ln[\tau_L/(\tau_0L^{\omega})]/L^{\psi}$ must display data collapse using $\psi=1/2$, $\omega=1$, and appropriately chosen constants $\epsilon_0$ and $\tau_0$. As shown in Fig.~\ref{fig_ed}, this is indeed the case, although the convergence to the limit distribution, as it is particularly noticeable for the effective multiplicity variable $\tau_L\epsilon$ (Fig.~\ref{fig_ed}b), is slow. 
\begin{figure*}[htbp!]   
    \centering
     \includegraphics[width=.32\linewidth]{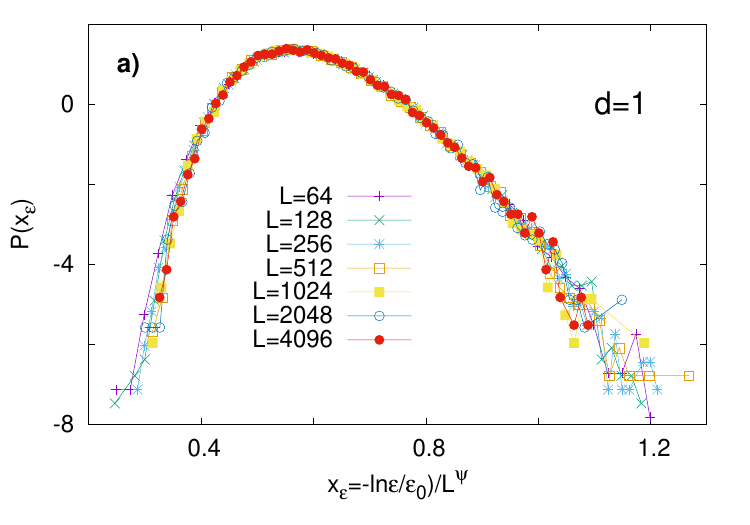}
     \includegraphics[width=.32\linewidth]{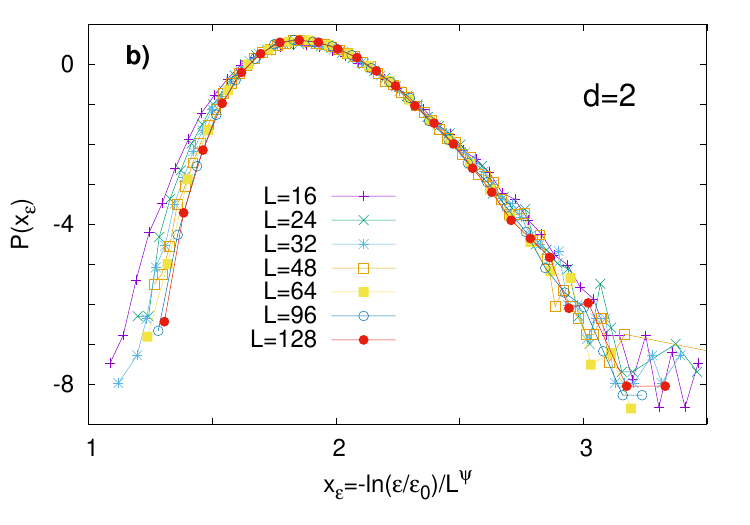}
     \includegraphics[width=.32\linewidth]{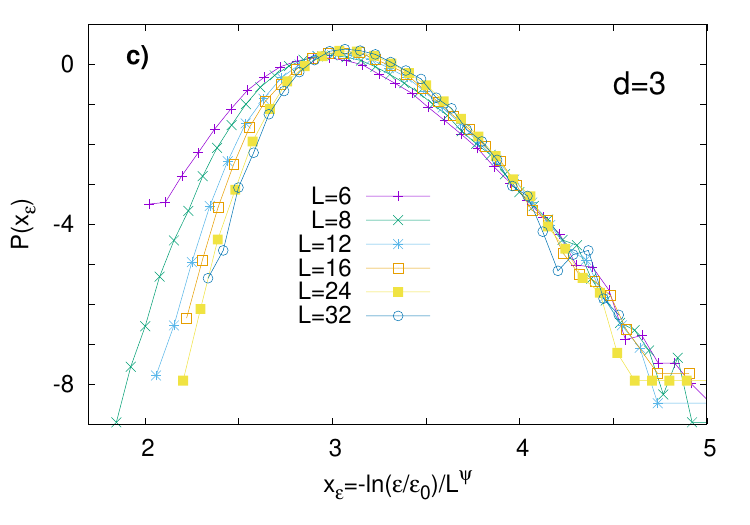}
     \includegraphics[width=.32\linewidth]{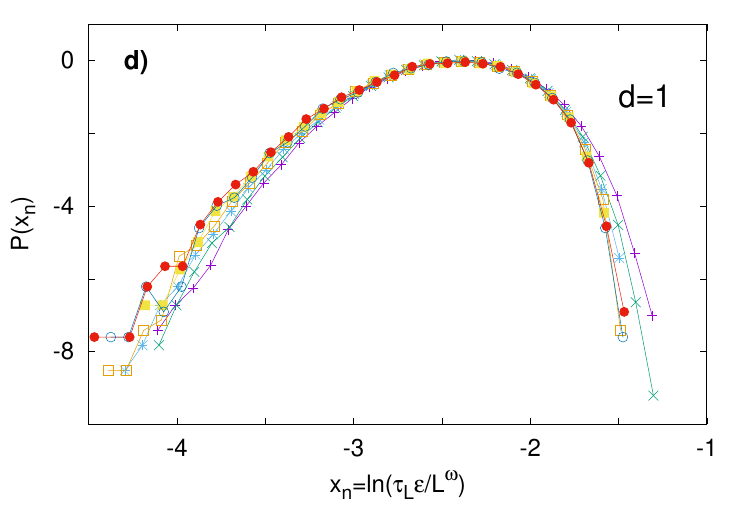}
      \includegraphics[width=.32\linewidth]{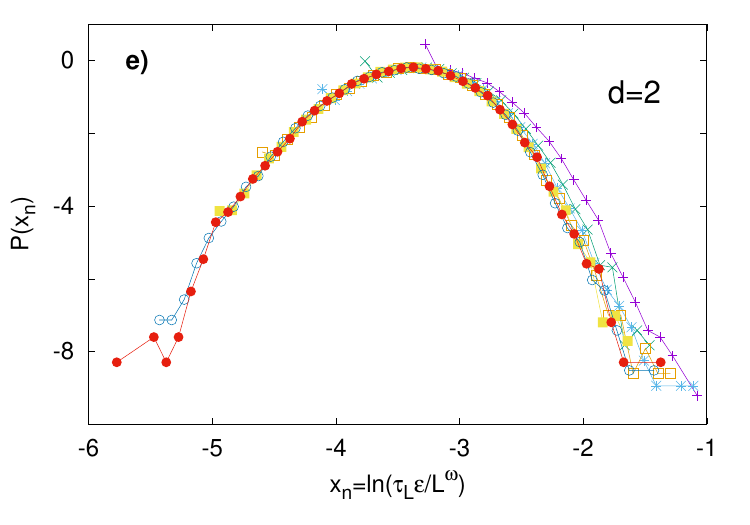}
      \includegraphics[width=.32\linewidth]{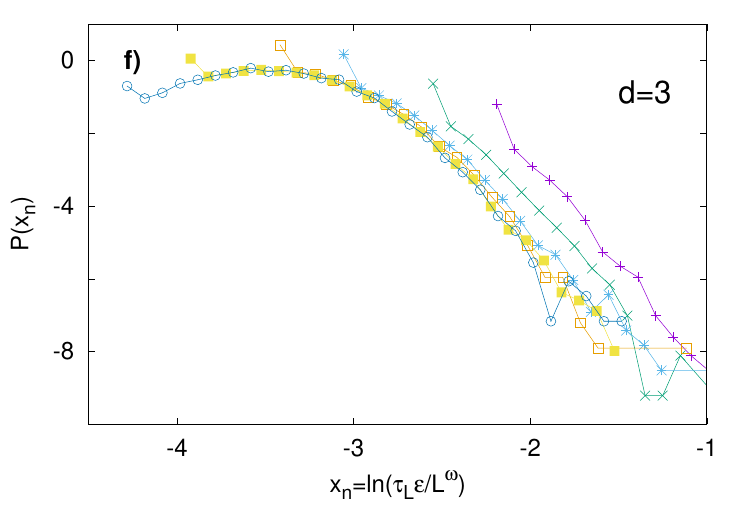}    
        \includegraphics[width=.32\linewidth]{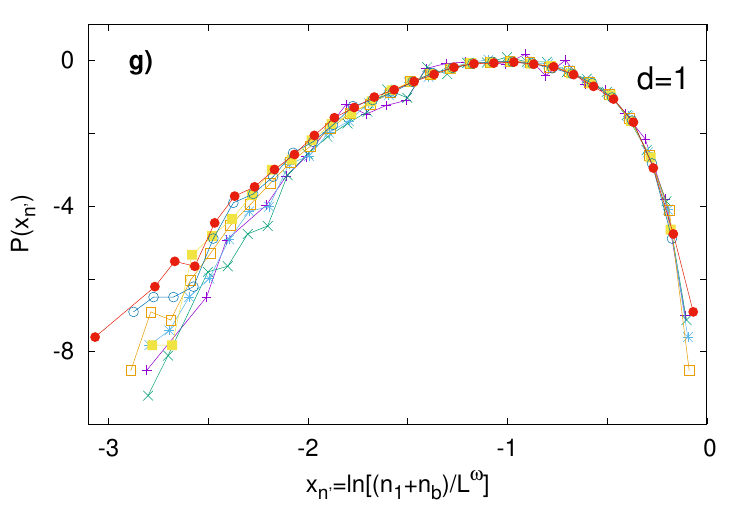}
      \includegraphics[width=.32\linewidth]{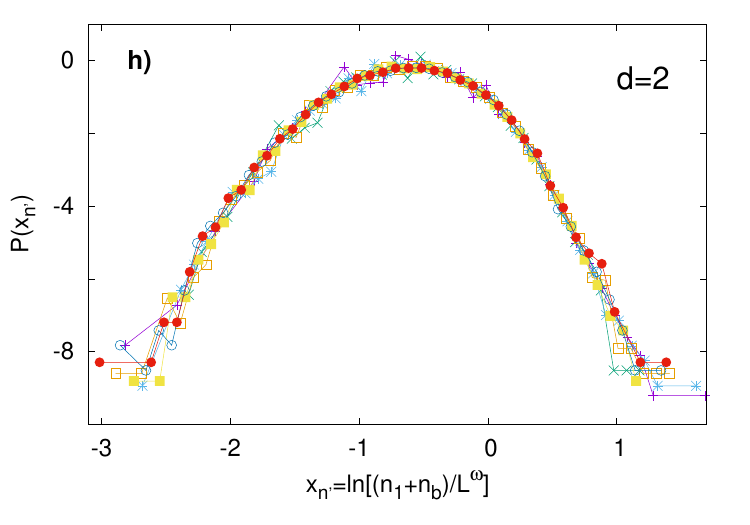}
      \includegraphics[width=.32\linewidth]{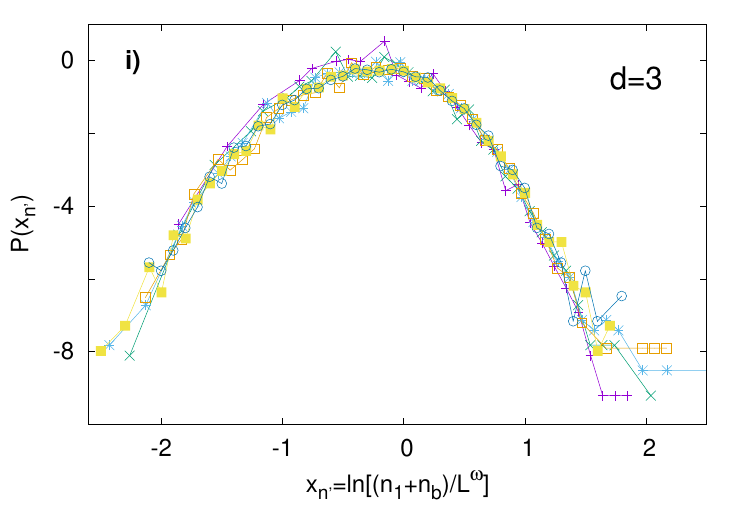}
       \includegraphics[width=.32\linewidth]{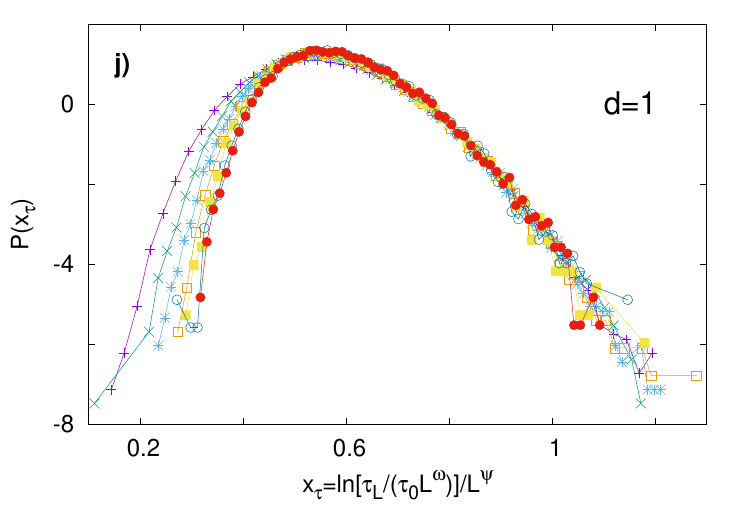}
     \includegraphics[width=.32\linewidth]{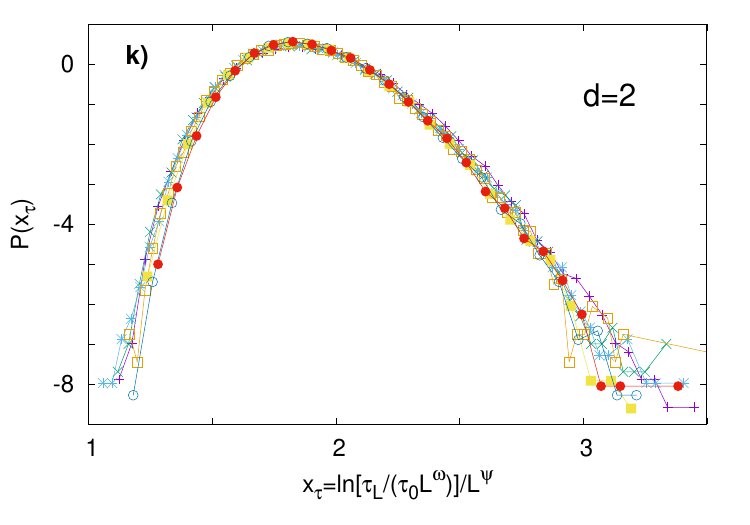}
      \includegraphics[width=.32\linewidth]{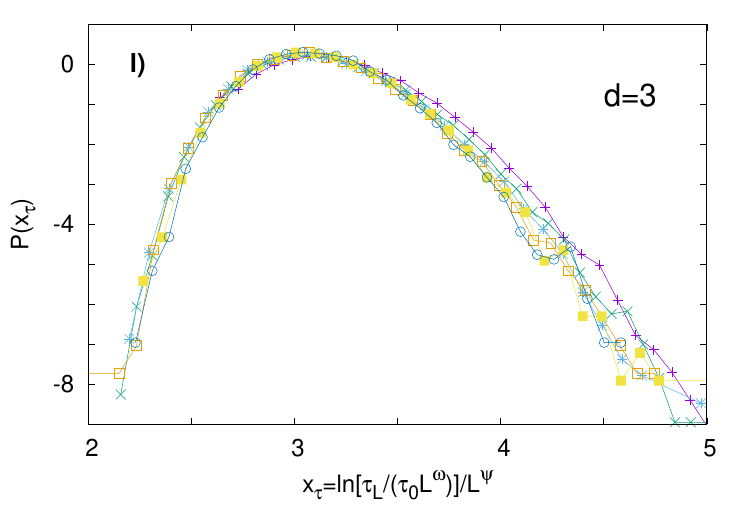}
     \vskip-5mm
       \caption{\textbf{Finite-size scaling of different properties of the dominant fusion obtained by the SDRG method.}
       \textbf{a)-c)} Distributions of the excitation energy in terms of $x_{\epsilon}=-\ln(\epsilon/\epsilon_0)/L^\psi$ a) in $d = 1$ using $\epsilon_0=4$, b) $d = 2$, using $\epsilon_0=33$ and c) $d = 3$ with $\epsilon_0=100$, aligning with the activated scaling relation in Eq.~(\ref{eps_L}), with the $\psi$ exponents of Table~\ref{table:1}.
       \textbf{d)-f)} Distributions of effective multiplicity $\tau_L\epsilon$  in terms of $x_n=\ln(\tau_L\epsilon/L^{\omega})$ d) for $d = 1$ using $\omega=1$, e) for $d = 2$ using $\omega=1.20(5)$, and f) for $d = 3$ using $\omega=1.25(5)$. 
       \textbf{g)-i)} Distributions of the multiplicity $n_1+n_b$ in terms of the variable $x_{n'}=\ln[(n_1 + n_b)/L^{\omega}]$ for different sizes in g) $d = 1$, using $\omega=1$, h) $d = 2$, using $\omega=1.25(5)$ and i) $d = 3$, using $\omega=1.4(1)$. 
       \textbf{j)-l)} 
       Distributions of $\tau_L$ in terms of $x=\ln[\tau_L/(\tau_0L^{\omega})]/L^\psi$, testing the activated scaling relation of Eq.~(\ref{tau_L}). j) in $d=1$ using $\tau_0=0.03$, k) in $d=2$, using $\tau_0=0.001$ and $\omega=1.25$, l) in $d=3$, using $\tau_0=0.0002$ and $\omega=1.4$.       
       } 
    \label{fig_sdrg_dist}
\end{figure*}
Thereafter, we computed the same distributions by the SDRG method. As it is demonstrated in the first column of panels of Fig.~\ref{fig_sdrg_dist}, these also fulfill the scaling relations quoted above, although with different constants $\epsilon_0$ and $\tau_0$.  

\begin{figure}[htbp!]   
    \centering
    \includegraphics[width=0.493\linewidth]{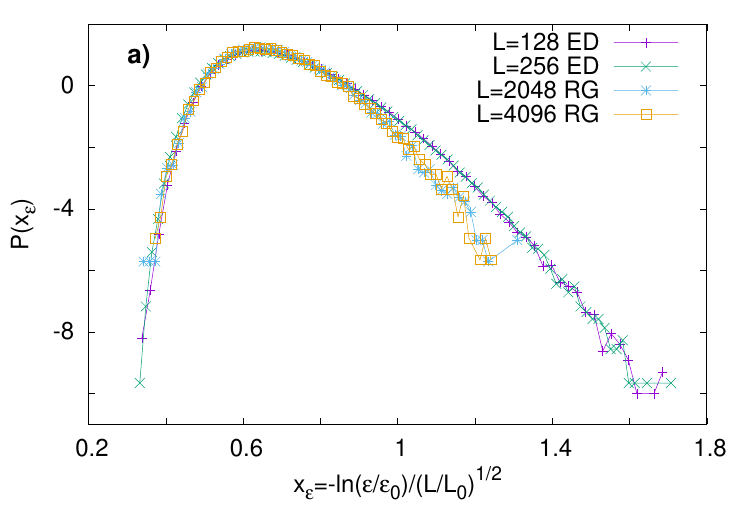}
    \includegraphics[width=0.493\linewidth]{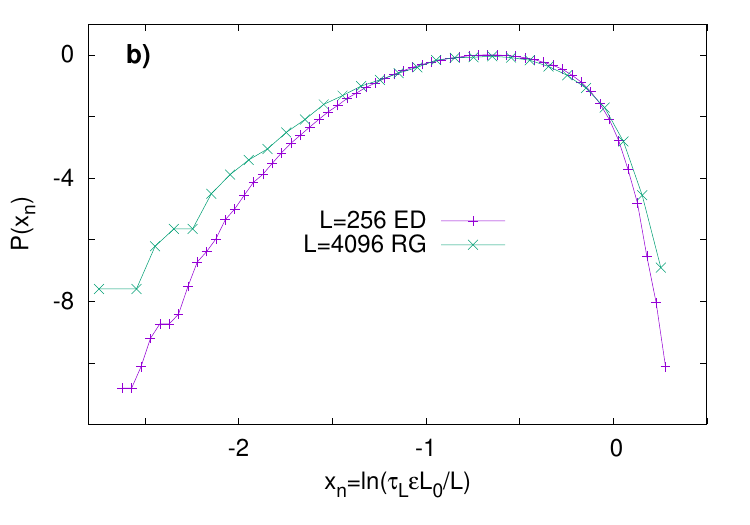}
    \includegraphics[width=0.493\linewidth]{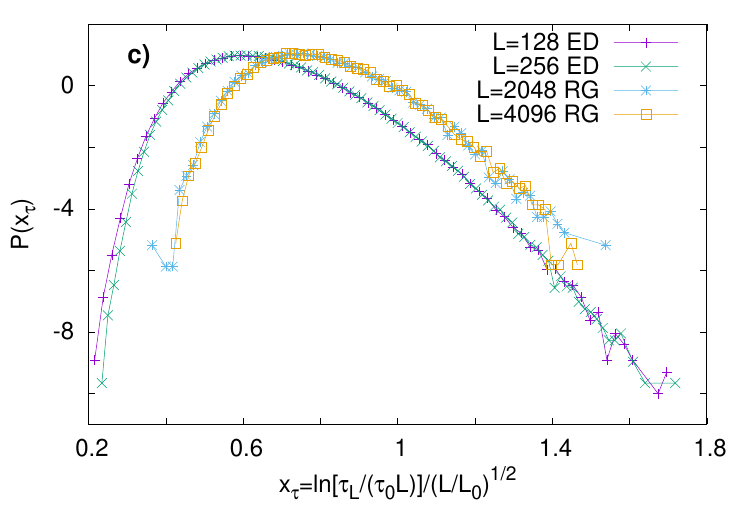}
    \includegraphics[width=0.493\linewidth]{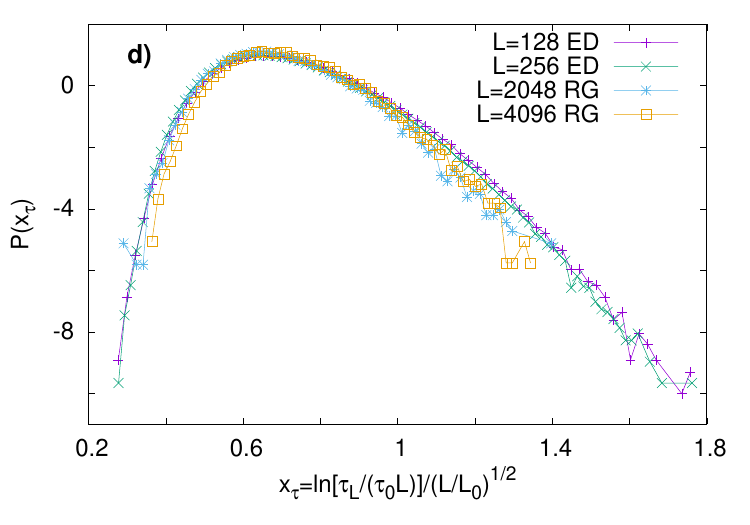}
    \vskip-5mm
     \caption{\textbf{Comparison of exact diagonalization and SDRG results in one dimension.} a) Gap distribution in terms of $x_{\epsilon}=-\ln(\epsilon/\epsilon_0)(L/L_0)^{-1/2}$, using parameters $\epsilon_0^{ED}=2$, $\epsilon_0^{RG}=4$, $L_0^{ED}=1$, and $L_0^{RG}=1.3$. b) Distribution of effective multiplicity in terms of $\ln(\tau_L\epsilon L_0/L)$, using parameters $L_0^{ED}=1$ and $L_0^{RG}=5.6$. c) and d) Distribution of threshold annealing time in terms of  $\ln[\tau_L/(\tau_0L)](L/L_0)^{-1/2}$ for two choices of fitting constants $\tau_0^{ED}=0.4$, $\tau_0^{RG}=0.03$, $L_0^{RG}=1.8$, and $L_0^{ED}=1$ (c), and $\tau_0^{ED}=0.2$, $\tau_0^{RG}=0.2$, $L_0^{RG}=1.6$, and $L_0^{ED}=1$ (d).}
     \label{fig_comp}
\end{figure}

The scaled distributions obtained by the two methods are compared in Fig.~\ref{fig_comp}. For the excitation energy, shown in Fig.~\ref{fig_comp}(a), the difference between the limit distributions can essentially be resolved by including a microscopic length scale $L_0$ in the scaling variable, $x_{\epsilon}=-\ln(\epsilon/\epsilon_0)(L/L_0)^{-1/2}$. This non-universal length scale is not expected to be reproduced correctly by the SDRG method. As demonstrated in Fig.~\ref{fig_comp}(b), such type of generalization of the scaling variable including $L_0$ results in a satisfactory collapse for the effective multiplicity, as well. 
Concerning the threshold annealing time $\tau_L$, the scaled distributions can be brought to the same shape by using a suitable $L_0$, see Fig.~\ref{fig_comp}(c), but a slight horizontal shift between the two master curves remains. Nevertheless, this difference can be greatly diminished by using parameters $\tau_0$ that differ from those giving an optimal collapse of data within each method [Fig.~\ref{fig_comp}(d), where $\tau_0=0.2$ is used for both methods]. This is a warning sign that the parameter $\tau_0$ found by an optimal collapse of limited-size data may be considerably different from its asymptotic value. Overall, we conclude that the SDRG method successfully identifies all universal aspects of the scaling behavior.

\subsection{Characteristics of the dominant fusion}

After validating the SDRG method in one dimension, we proceed with presenting detailed results obtained by the SDRG method in higher dimensions.  
We begin with showing different characteristics of the dominant fusion event which determines the threshold annealing time $\tau_L$ of a finite sample of size $L$.
Among these characteristics, we consider first the excitation energy $\epsilon$ of the dominant fusion. The distributions of $\epsilon$ obtained by the SDRG method for different sizes in dimensions $d=1,2,$ and $3$ are shown in Fig.~\ref{fig_sdrg_dist}(a)--(c), respectively.
As can be seen in the figure, the scaling plots of distributions are in satisfactory agreement with the relation in Eq.~(\ref{eps_L}), using estimates of the tunneling exponent $\psi$ from earlier works~\cite{fisher,kovacs2d,kovacs3d}. Nevertheless, in higher dimensions, where the numerically accessible system sizes are more limited, the scaling collapse of the data is poorer, as it can be expected.     

Next, we consider the effective multiplicity $\tau_L\epsilon$ of the dominant fusion, the distributions of which are shown in Fig.~\ref{fig_sdrg_dist}(d)--(f).
These are found to follow the algebraic scaling anticipated in Eq.~(\ref{n_L}) in all dimensions, with exponents $\omega=1$ in $d=1$, $\omega=1.20(5)$ in $d=2$, and $\omega=1.25(5)$ in $d=3$. These values are not far from the exponents characterizing the scaling of the size of the energy cluster, listed in Table~\ref{table:1}. 
By the SDRG method, it is also possible to directly calculate the multiplicity $n_1+n_b$ of the dominant fusion. The distributions, shown in Fig.~\ref{fig_sdrg_dist}(g)--(i), follow the expected algebraic scaling with estimates $\omega=1.25(5)$ in $d=2$
 and $\omega=1.4(1)$ in $d=3$ that are very close to the exponents of the energy cluster quoted in Table~\ref{table:1}.

Finally, we considered the sample-to-sample distribution of the threshold annealing time $\tau_L$, which is the most relevant quantity from the perspective of QA. As it is shown in Fig.~\ref{fig_sdrg_dist}(j)--(l), the scaling plots are compatible with the activated finite-size scaling of $\tau_L$ in Eq.~(\ref{tau_L}).  

\subsection{Characteristics of all fusions}

\begin{figure*}[htbp!]   
    \centering
    \includegraphics[width=.96\linewidth]{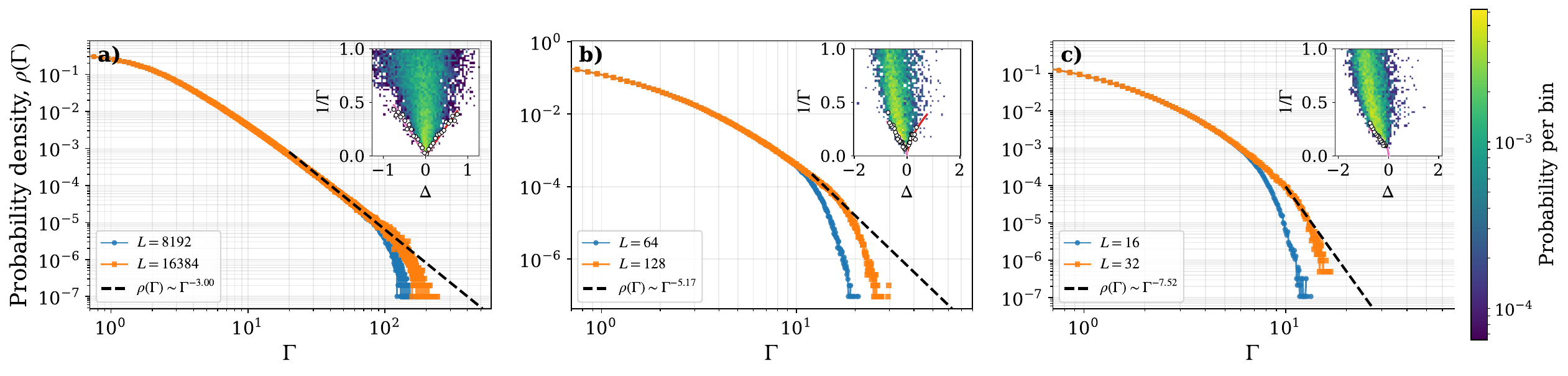}
\vskip-4mm
    \caption{\textbf{Distribution of excitation energies in one, two, and three dimensions.} Panels a), b), and c) show the probability density of $\Gamma=-\ln\epsilon$ of all fusion events, pooled over many independent disorder realizations of a given size, with approximately $10^8$ fusion events in total for each size ($2\times10^7$ for $L=32$ in $d=2$, due to computational constraints), for the system sizes indicated in each panel, in $d =1$, $d=2$, and $d=3$, respectively. Dashed lines show the predicted power law $\rho(\Gamma)\sim\Gamma^{-\kappa}$ [Eq.~(\ref{tail})] with $\kappa=1+d/\psi$ [Eq.~(\ref{kappa2})], giving exponents $-3.00$, $-5.17$, and $-7.52$. Insets: joint probability distribution of $1/\Gamma$ and the reduced control parameter $\Delta$ of fusions, for the largest system size studied in each dimension ($d = 1$: $L = 131072$; $d = 2$: $L = 362$; $d = 3$: $L = 40$, pooled over 5 independent realizations to compensate for otherwise too-sparse statistics). Red and magenta curves show the fits $1/\Gamma = A_{\pm} |\Delta|^{\psi\nu}$ [Eq.~(\ref{eq:envelope})] to the $\Delta>0$ and $\Delta<0$ branches; open circles mark the 1st-percentile points used in the fit. In $d = 3$, the $\Delta > 0$ branch has too few near-critical boundary points for a reliable fit and is omitted.}    
    \label{fig_eps_all}
\end{figure*}

\begin{figure*}[htbp!]   
    \centering
    \includegraphics[width=.96\linewidth]{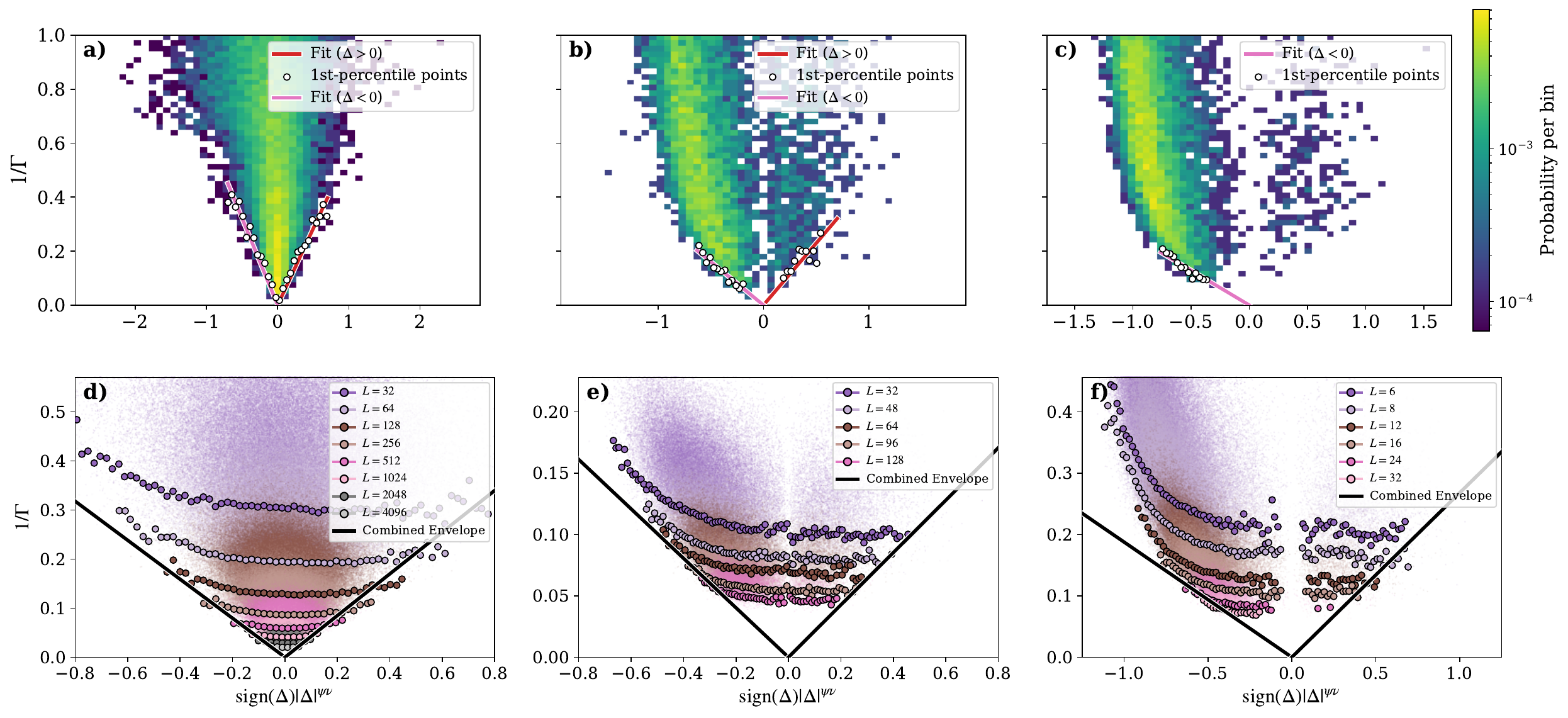}
\vskip-4mm
    \caption{\textbf{Scaled distribution of fusion events in one, two, and three dimensions.} Panels a)--c) show the joint probability distribution of $1/\Gamma$ and the scaled control parameter $\text{sign}(\Delta)|\Delta|^{\psi\nu}$, built from all fusion events of the same large samples as in the insets of Fig.~\ref{fig_eps_all}. 
    Open circles mark the first-percentile points, while red and magenta lines show linear fits on each side of $\Delta=0$. In $d=3$, the $\Delta>0$ branch is omitted for lack of data. 
    In panels d)--f), using the same variables, data points represent the dominant fusion event (the one with maximal $\tau_c$) of sets of samples with different size. 
    Solid circles trace the lower 5th-percentile envelope of $1/\Gamma$, separately for each $L$. The black straight lines are guides to the eye.
    }    
    \label{fig:max_tau_and_heat_map}
\end{figure*}

In this section, we examine the distributions of different properties associated with all fusion events of a single (large) sample. As the same distributions are obtained when the statistics of all fusion events are aggregated over many independent samples of a given size, the marginal distributions shown below are computed in this way, as it yields better statistics in the tails.
We start with the joint probability density of the excitation energy $\epsilon$ and the reduced control parameter $\Delta=\theta-\theta_c$ of fusions. As $\epsilon$ varies through orders of magnitude, it is expedient to plot the joint probability density $g(1/\Gamma,\Delta)$ with the variable $1/\Gamma=-1/\ln\epsilon$ instead of $\epsilon$. 
This is shown in the insets of Fig.~\ref{fig_eps_all}(a)--(c) in dimensions $d=1,2$, and $3$, respectively, for the largest sample used for the statistics of all fusions. 
As can be seen in the insets, in $d=1$ [panel (a)] the control parameters of fusions are distributed largely symmetrically around the critical point $\theta_c$ even at high energies, due to the self-duality of the model~\cite{fisher}. For $d>1$ [panels (b) and (c)], in contrast, there is a considerable shift towards the ferromagnetic phase, which is diminishing with decreasing energy.  
The envelope of the domain containing an $O(1)$ fraction of excitations is visualized in the figures by the first percentiles of conditional distribution of $1/\Gamma$ at a series of control parameter values in the vicinity of the critical point. The expected form of the envelope at the cusp, $1/\Gamma_{\Delta}\sim |\Delta|^{\psi\nu}$, 
see Eq.~(\ref{eq:envelope}), is shown by the red and magenta curves, which are fitted independently on the two sides of $\Delta=0$ to the first-percentile points. The envelope is also illustrated in a linearized plot in Fig.~\ref{fig:max_tau_and_heat_map}(a)--(c), where the joint probability density $g(1/\Gamma,{\rm sign}(\Delta)|\Delta|^{\psi\nu})$ and the corresponding first percentiles are plotted. 
In panels (a)--(c), the first percentiles marking out the envelope indeed approach the origin linearly. For $d=2$ and $d=3$, however, the data points remain relatively far from the origin, owing to the limited system sizes accessible numerically.  
In panels (d)--(f) of Fig.~\ref{fig:max_tau_and_heat_map}, the same variables, $1/\Gamma$ and ${\rm sign}(\Delta)|\Delta|^{\psi\nu}$ are used, but, here, the points represent data of the dominant fusions for sets of samples with different sizes. Plotted in this way, the cloud of data points shrinks to the origin proportionally to $L^{-\psi}$ in both coordinates, i.e.~along a linear path. 

The marginal distributions of the logarithmic excitation energies $\Gamma=-\ln\epsilon$ of all fusions are shown in Fig.~\ref{fig_eps_all}. The bulk of the distribution is independent of the system size, while finite-size effects appear as a cutoff at low energies (large $\Gamma$) shifting proportionally to $L^{\psi}$ as implied by Eq.~(\ref{eps_L}). 
According to our scaling results, the tail must be algebraic, $\rho(\Gamma)\sim\Gamma^{-\kappa}$ with $\kappa=1+d/\psi$ [Eq.~(\ref{kappa2})] in all three dimensions. 
In $d=1$, the power law with $\kappa=3$ is clearly visible before the cutoff, whereas, in $d=2$ and $3$, the asymptotic domain seems to begin at or beyond the cutoff determined by the computationally accessible system sizes, preventing us from an independent confirmation of Eq.~(\ref{kappa2}).

\begin{figure*}[htbp!]      
    \centering
    \includegraphics[width=.96\linewidth]{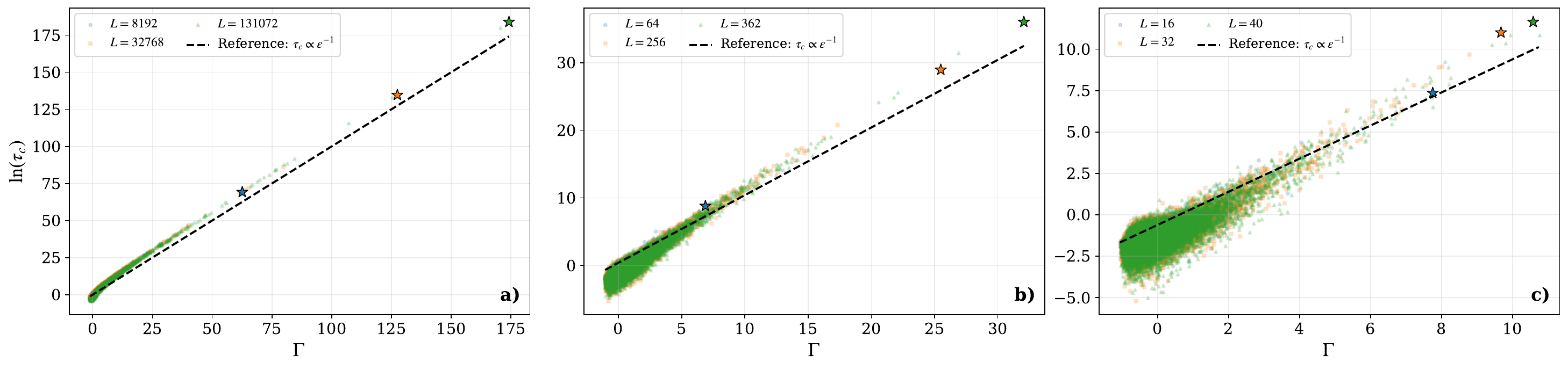}
    \vskip-4mm
    \caption{\textbf{Threshold annealing time vs.~energy gap at fusion events.} Panels a), b), and c) show $\ln(\tau_c)$ vs.\ $\Gamma$ for all fusion events, for system sizes indicated in each panel in $d = 1$, $d = 2$, and $d = 3$, respectively. Stars mark the dominant fusion event having the maximal $\tau_c$ in each sample. The dashed line indicates the reference scaling $\tau_c\propto\epsilon^{-1}$.}
    \label{fig:tau_vs_eps}  
\end{figure*}

Next, we turn to the crossover time $\tau_c$ of all fusion events. 
Writing the relationship between $\tau_c$ and the energy in Eq.~(\ref{taueps}) in terms of logarithmic variables, the multiplicity in the numerator of Eq.~(\ref{taueps}) will give only a correction to the leading order behavior $\ln\tau_c\sim \Gamma$. 
Indeed, we find data of individual events shown in Fig.~\ref{fig:tau_vs_eps} 
in one, two, and three dimensions to broadly follow this asymptotically linear relationship indicated by the dashed reference line.

\begin{figure*}[htbp!]   
    \centering
    \includegraphics[width=.96\linewidth]{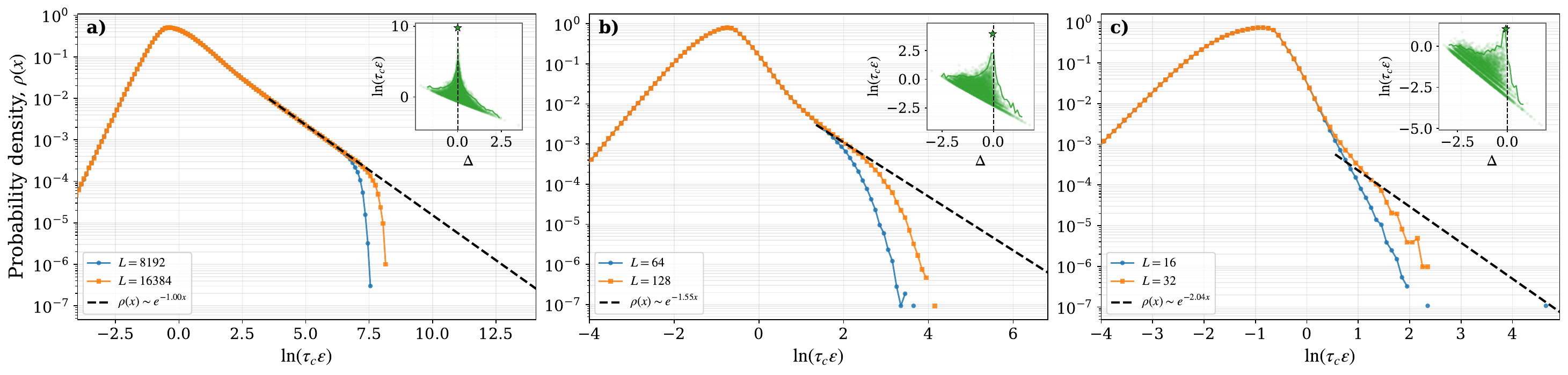}
\vskip-4mm
    \caption{\textbf{Probability density of $\ln(\tau_c\epsilon)$ in one, two, and three dimensions.} Panels a), b), and c) show the probability density $\rho(x)$ of $x=\ln(\tau_c\epsilon)$ for the same fusion events as in Fig.~\ref{fig_eps_all}, in $d = 1$, $d = 2$, and $d = 3$, respectively. The expected tail, $\rho(x)\sim e^{-(d/\omega)x}$, see the text, is indicated by the dashed lines with $d/\omega=1.00$, $1.55$, and $2.04$. Insets: Data points with coordinates $\Delta$ and $\ln(\tau_c\epsilon)$ represent all fusion events of a single sample with $L = 131072$ in $d = 1$, $L = 362$ in $d = 2$, and $L = 40$ in $d = 3$. The solid curves show the binned 99th percentile, the star marks the event with the largest $\tau_c$, and the vertical dashed line denotes the critical point. 
    }    
    \label{fig:ln_tau_c_eps_with_inset}
\end{figure*}

Concerning the correction to this leading-order relationship, the distribution of the logarithm of the effective multiplicity, $\ln(\tau_c\epsilon)$, for all fusion events is shown in Fig.~\ref{fig:ln_tau_c_eps_with_inset}, with its dependence on $\Delta$ in the insets. 

As for $\Gamma$ sampled for all fusions, the bulk of distributions is independent of the system size and the finite-size effect manifests itself in a cutoff at the tail, shifting proportionally to $\omega\ln L$.   
Since $\tau_c\epsilon\sim n_1+n_b$ by Eq.~(\ref{taueps}), the variable $x=\ln(\tau_c\epsilon)$ is, up to an additive constant, the logarithm of the multiplicity of the fusion. Thus, the algebraic tail in Eq.~(\ref{mtail}) transforms into an exponential one, $\rho(x)\sim e^{-(d/\omega)x}$. 
For $d=1$, the numerical distribution is in a good agreement with this prediction while, for $d=2$, the tendency to the asymptotic law is just emerging at the cutoff region. Finally, for $d=3$, the asymptotic region must be well beyond the observable domain limited by finite-size cutoff. 

\begin{figure*}[htbp!]   
    \centering
    \includegraphics[width=.96\linewidth]{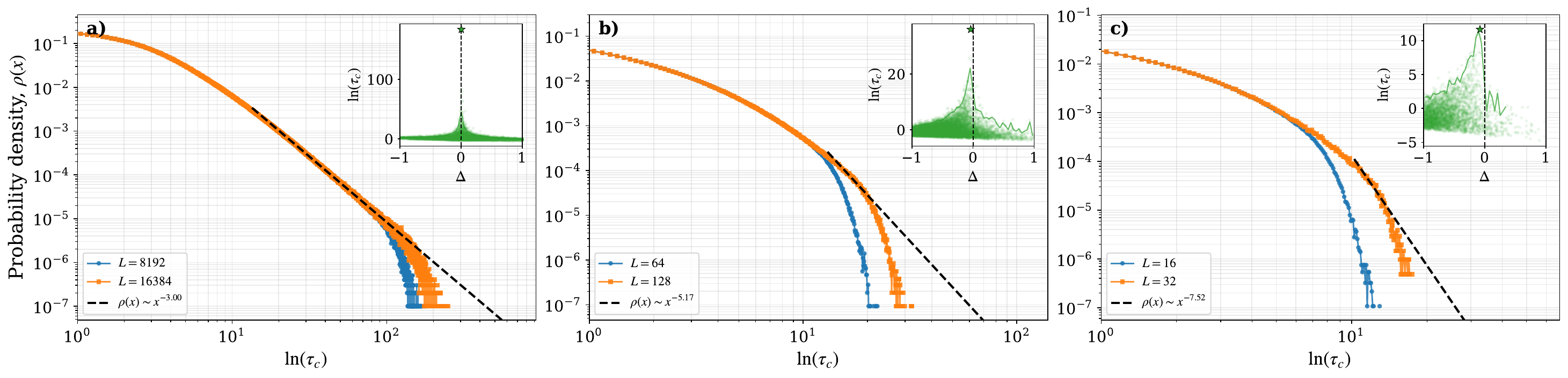}
\vskip-4mm
    \caption{\textbf{Probability density of the threshold annealing time $\ln(\tau_c)$ in one, two, and three dimensions.} Panels a), b), and c) show the probability density $\rho(x)$ of $x=\ln(\tau_c)$ for the same fusion events as in Fig.~\ref{fig_eps_all}, in $d = 1$, $d = 2$, and $d = 3$, respectively. The variable $\ln\tau_c$ is expected to share the algebraic tail of $\Gamma$ [Eqs.~(\ref{tail}) and (\ref{kappa2})], $\rho(x)\sim x^{-(1+d/\psi)}$, shown by the dashed lines with exponents $-3.00$, $-5.17$, and $-7.52$, computed from the values of $\psi$ in Table~\ref{table:1}. Insets: $\ln(\tau_c)$ vs.~the reduced control parameter $\Delta$ for all fusion events of the same sample as in the insets of Fig.~\ref{fig:ln_tau_c_eps_with_inset}. 
    The solid curves show the binned 99th percentile of $\ln(\tau_c)$, the star marks the event with the largest $\tau_c$, and the vertical dashed line denotes the critical point.}    
    \label{fig:ln_tau_c_with_inset}
\end{figure*}

The distribution of $\ln(\tau_c)$ over all fusion events is shown in Fig.~\ref{fig:ln_tau_c_with_inset}, together with its dependence on the control parameter in the inset. 
The tail of the probability density of $x=\ln\tau_c$ in an infinite system is expected to have the form $\rho(x)\sim x^{-1-d/\psi}$, and, as for the previous two characteristics of fusions, the data follow this form up to the finite-size cutoff at large $\tau_c$.

\begin{figure*}[htbp!]   
    \centering
    \includegraphics[width=.96\linewidth]{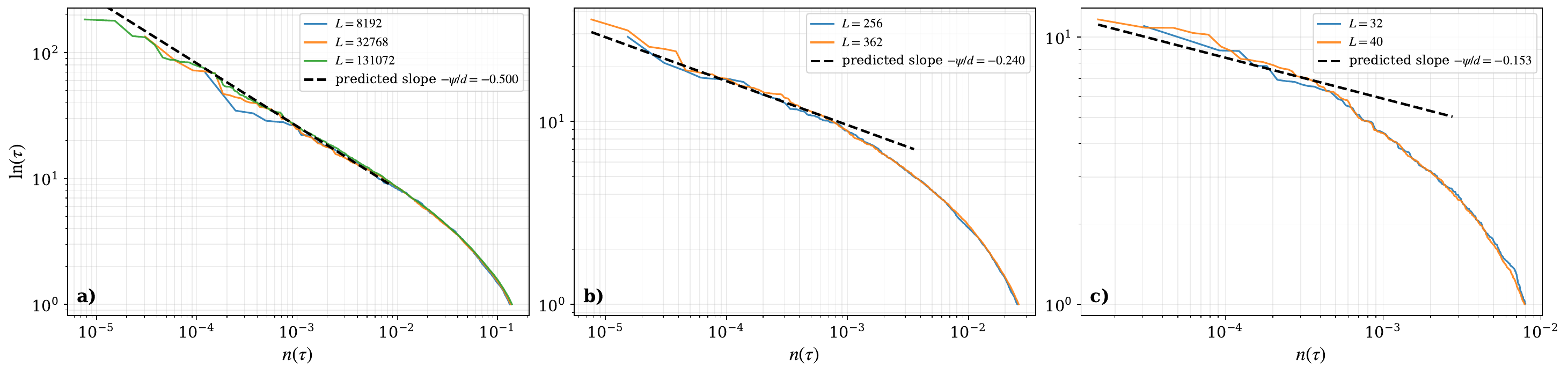}
\vskip-4mm
    \caption{\textbf{Scaling of the error density with annealing time.}     Panels a), b), and c) show $\ln(\tau)$ against $n(\tau)$, the fraction of fusion events with $\tau_c>\tau$, on a log-log plot for $d=1$, $d=2$, and $d=3$, respectively. The predicted asymptotic dependence is indicated by the dashed lines having slopes $-\psi/d$.}
    
    \label{fig:tau_c_vs_n_leading_order}
\end{figure*}

Finally, we calculated the $\tau$-dependence of error density $n(\tau)$ which is measured as the fraction of fusions having a threshold annealing time $\tau_c$ greater than $\tau$. In order to test the leading-order scaling in Eq.~(\ref{eq:tau_c_n_scaling}), $\ln\tau\sim[n(\tau)]^{-\psi/d}$, valid in the large-$\tau$ limit, we plotted $\ln\tau$ against $n(\tau)$ on a double logarithmic scale in Fig.~\ref{fig:tau_c_vs_n_leading_order}.
In this plot, the graph of the function for large $\tau$ is expected to be linear with a slope $-\psi/d$, which is in accordance with the measured slopes in all three dimensions.

\section{Discussion}
\label{sec:discussion}

In the quantum annealing of the RTFIM, the strong disorder present in the system leads to an effective decomposition of the complex process of error formation into a series of distinct elementary events. 
This picture emerges in the framework of the SDRG approach that represents the annealing process as a gradual aggregation of strongly coupled spin clusters having a highly non-trivial, spatially disconnected structure~\cite{kovacs2020-contactprocess}. 
The instantaneous low-energy excitations associated with defect formation are identified in this picture as fusions (or microscopic paramagnetic to ferromagnetic transitions) between pairs of strongly coupled clusters. 
Thereby, this approach goes beyond a cumulative characterization of the annealing process, making accessible the individual elements of defect formation. 
Based on this approach, we studied in this work the statistics of different properties of all fusion events of a large sample, as well as properties of the bottleneck fusion event of samples in dimensions $d=1,2$, and $3$. 

Key open questions in higher dimensions include the scaling of the bottleneck annealing rate with the system size, as well as the asymptotic dependence of the defect density on the annealing rate. Here, we addressed these questions by formulating a scaling theory valid in arbitrary dimensions, and which is in accordance with analytical SDRG results obtained earlier for the one-dimensional model~\cite{juhasz2025}. 
Our results are confirmed by large-scale numerical calculations relying on the SDRG technique. To achieve these results, we developed a scheme to perform annealing efficiently and when possible, narrowed down the calculations to the relevant critical region of the annealing process. 
The SDRG method was validated for the one-dimensional model by a comparison with results of exact diagonalization based on the free-fermion technique.  

According to our results, the bottleneck annealing time needed to avoid defects in a finite sample scales with the linear size $L$ as $\tau_L\sim L^{\omega}e^{cL^{\psi}}$ in all studied dimensions. 
This relationship corresponds to an ultra-slow decrease of the defect density with the annealing time, as $n(\tau)\sim (\ln\tau)^{-d/\psi}$ in leading order for large $\tau$.  
We pointed out that, besides the excitation energy of a fusion, also the multiplicity, i.e.~the number of original fields involved in the fusion event, affects the crossover time scale of the fusion. The latter can also be interpreted as each fusion having its own renormalized annealing rate, differing from the ``bare'' annealing rate $\tau^{-1}$ by a factor proportional to the multiplicity. 
This circumstance leads to the appearance of the factor $L^{\omega}$ in the scaling form of $\tau_L$, containing an additional non-trivial critical exponent, the multiplicity exponent $\omega$.
By two different scaling arguments, we find this exponent to be consistently related to the usual critical exponents as $\omega=\psi+1/\nu$. 

The numerical analysis presented in this work has been performed for a particular form of disorder and annealing protocol. Nevertheless, it is well-known that the critical exponents of an IDFP are universal in the sense that they are independent from the particular form of disorder~\cite{fisher, Motrunich2000, kovacs2d, kovacs3d}. Consequently, our asymptotic results valid in the limit of long annealing times or large system sizes, are expected to be independent from the type of disorder.  
Concerning the other generality-breaking factor of our model, the time-dependence of the control parameter, we note that the region where error formation is considerable shrinks to the critical point with increasing  annealing time. As a consequence, our asymptotic results are expected to be  universally valid for protocols having a linear time-dependence at the critical point.

Even with our best efforts, some of the results are not yet probing the asymptotic regime in $d=3$ dimensions, as the tail of corresponding distributions is dominated by rare events that would require much larger sizes to sample effectively.
Possible alternatives could be the current experiments carried out with programmable superconducting annealers. At the present time, these consist of a few thousand qubits~\cite{king,king2023_b}, which match the sizes of two-dimensional systems studied in this work.

Even though we only considered the ferromagnetic model in this paper, our results can serve as important starting points toward understanding the practically more relevant situation of spin glasses that contain some fraction of antiferromagnetic couplings, as well. Indeed, for strong enough disorder, the asymptotic scaling behavior in both cases is widely expected to be governed by the same IDFP~\cite{Motrunich2000}. Nevertheless, a recent quantum Monte Carlo study performed for a two-dimensional Ising spin glass with a maximal linear size $L=24$ concluded that the parity-restricted gap relevant for QA closes algebraically with the system size rather than exponentially~\cite{Bernaschi2024}.

Although we considered in this paper the RTFIM on regular lattices, the picture of gradual aggregation of strongly coupled clusters is valid more generally, even for irregular lattices or networks~\cite{Juhasz_2013}, provided the phase transition on the particular structure is appropriately described by an IDFP. 
Such an example is the RTFIM on a site-diluted lattice~\cite{senthil1996}. 
Having only nearest-neighbor interactions as before results in the fragmentation of the system into independent components. From the perspective of QA, it is reasonable to restrict the system to the largest connected component of the diluted lattice. Above the percolation threshold, this component is compact, and the phase transition on this structure is known to be governed by the same IDFP as for coupling-disorder on the corresponding non-diluted lattice.  This, however, changes at the percolation threshold where the largest component is a fractal. Here, the critical point of the RTFIM, termed as multicritical point, is still described by an IDFP, however, with a set of critical exponents different from that of the coupling-disorder case~\cite{Kovacs2022-qmcp, Kovacs2024-qmcp, Love2025,luzzatto2025}. 
We expect our scaling theory of QA presented in this work to be valid also in this case, with the appropriate critical exponents of the multicritical point.

Another important future direction is to explore if the limiting annealing rate can be substantially increased by appropriately designing the local fields of the system. A prominent direction is to choose the fields in a correlated way that lowers the strength of the effective disorder to the extent that conventional scaling replaces the ultra-slow, activated dynamics of the IDFP. In one-dimensional systems, it has been shown that there are in principle a continuum of ways to achieve this~\cite{hoyos,Shirai2021,juhasz2022,roosz2022,knysh}, awaiting to be generalized to higher dimensional systems.

\begin{acknowledgments}
This work was supported by the National Research, Development and Innovation Office NKFIH under Grant No.~K146736. The work of I.A.K.~and N.S.~was supported by the National Science Foundation under Grant No.~PHY-2310706 of the QIS program in the Division of Physics. 
\end{acknowledgments}




\bibliography{references}

\appendix

\section{Characteristics of cluster fusion}
\label{maximum}

In this section, we consider a single fusion event between two clusters, treated isolated from the rest of the system. 
The effective Hamiltonian of the subsystem is
\be
\mathcal{H}_{\rm 12}(t)=-\tilde J(t)\tilde\sigma_1^x\tilde\sigma_2^x-\tilde h_1(t)\tilde\sigma_1^z -\tilde h_2(t)\tilde\sigma_2^z.
\label{H_eff}
\ee
As the interaction with the rest of the system is negligible, the state of the subsystem remains in the even sector spanned by the eigenvectors of $\mathcal{H}_{\rm 12}(t)$
\beqn
|E_-\rangle=\frac{1}{\sqrt{1+\gamma^2}}(|\uparrow\uparrow\rangle+\gamma |\downarrow\downarrow\rangle) \nonumber \\
|E_+\rangle=\frac{1}{\sqrt{1+\gamma^2}}(\gamma|\uparrow\uparrow\rangle-|\downarrow\downarrow\rangle).
\eeqn
where $\gamma(t)=\sqrt{1+r^2(t)}-r(t)$ and
$
r(t)\equiv[\tilde h_1(t)+\tilde h_2(t)]/\tilde J(t)
$. 
The corresponding energies are
$E_{\mp}=\mp\tilde J(t)\sqrt{1+r^2(t)}$. 
Since the effective bond (fields) monotonically increases (decrease) with $t$, there must be a point $t_0$ at which $\tilde J(t_0)=\tilde h_1(t_0)$. This point approximately coincides with the minimum position of the energy gap $\epsilon(t)=E_+(t)-E_-(t)$, the difference becoming irrelevant at large length scales~\cite{juhasz2025}. 

According to the adiabatic theorem, the system will remain in the ground state of the instantaneous Hamiltonian as long as the annealing time satisfies
\be
\tau\gg\max_{0\le s\le 1}\frac{\left|\langle E_+(s)|\frac{d{\cal H}_{12}(s)}{ds}|E_-(s)\rangle\right|}{\epsilon^2(s)},
\label{at}
\ee
where $s\equiv t/\tau$ is the dimensionless time introduced in Sec.~\ref{sub:model}, such that the original transverse fields decrease as $h_i(s)=(1-s)g_0h_i^0$, see Eq.~(\ref{hit}).
The effective parameters can be written as 
\beqn
\tilde J(s)&=&J^0(1-s)^{-n_b}, \nonumber \\
\tilde h_i(s)&=&h_i^0(1-s)^{n_i}, \quad i=1,2,
\eeqn
where $J^0$, $h_1^0$, and $h_2^0$ do not depend on $s$, whereas $n_b$, $n_1$, and $n_2$ denote the number of times the original field appears in the product form of the effective parameters.

The matrix element in the numerator of the adiabatic theorem can be evaluated to give
\be
\left|\frac{r}{\sqrt{1+r^2}}\frac{d\tilde J}{ds}-\frac{1}{\sqrt{1+r^2}}\frac{d}{ds}(\tilde h_1+\tilde h_2)\right|.
\ee
Using the time dependence of effective parameters, the matrix element becomes
\be
\left(\frac{r}{\sqrt{1+r^2}}n_b\tilde J+\frac{1}{\sqrt{1+r^2}}(n_1\tilde h_1+n_2\tilde h_2)\right)\frac{1}{1-s},
\ee
while the energy gap is $\epsilon(s)=2\tilde J(s)\sqrt{1+r^2}$.
The ratio in the adiabatic theorem, $Q(s)$, equal to the matrix element divided by the squared gap $\epsilon^2(s)$, can be written as
\beqn
Q(s)=\frac{1}{4[1+r^2(s)]^{3/2}\tilde J(s)} \nonumber \\
\times \frac{\tilde h_1(s)(n_b+n_1)+\tilde h_2(s)(n_b+n_2)}{\tilde J(s) (1-s)}.
\eeqn
We are interested in the crossover time $\tau_c\equiv \max_{0\le s\le 1} Q(s)$. 
The function $Q(s)$ is complicated enough so that the maximum position $s^*$ cannot be calculated analytically.

We can, however, make use that, at low energy scales the two fields typically differ by orders of magnitude, i.e.~$\tilde h_1\gg \tilde h_2$.
 Therefore, in the rest of the section, we set $\tilde h _2=0$, which makes the calculations simpler.  By this approximation, we have
\be
Q(s)=\frac{r(s)}{4[1+r^2(s)]^{3/2}\tilde J(s)}\frac{n_b+n_1}{1-s}.
\ee
Searching for its maximum, $\frac{dQ}{ds}=0$ can be solved for the ratio $r(s)$ at the maximum position $s^*$:
\be
r^*\equiv r(s^*)=\sqrt{\frac{2n_b+n_1-1}{n_b+2n_1+1}}.
\ee
Note that this makes sense if $2n_b+n_1-1>0$. 
Using that
\beqn
\tilde J(s)&=&\tilde J(s_0)\left(\frac{1-s}{1-s_0}\right)^{-n_b}, \nonumber \\
\tilde h_1(s)&=&\tilde h_1(s_0)\left(\frac{1-s}{1-s_0}\right)^{n_1},
\eeqn
where $s_0\equiv t_0/\tau$, 
we can write
\be
r(s)=\left(\frac{1-s}{1-s_0}\right)^{n_1+n_b}.
\ee
Using this, we obtain ultimately for the maximum of $Q(s)$:
\beqn \label{eq: tau_c_Q_s_star}
\tau_c=Q(s^*)=\frac{1}{4\tilde J(s_0)}\left(\frac{2n_b+n_1-1}{n_b+2n_1+1}\right)^{\frac{2n_b+n_1-1}{2(n_1+n_b)}} \nonumber \\
\times \left(3\frac{n_1+n_b}{n_b+2n_1+1}\right)^{-3/2}\frac{n_1+n_b}{1-s_0}.
\eeqn
This formula is valid only for $2n_b+n_1-1>0$. The only invalid case is $n_b=0$, $n_1=1$, for which the maximum would be at $s=1$. Considering long annealing times, this case can be ignored since the energy scale of such a fusion event is in the order of magnitude of unrenormalized parameters.  

 The energy scale of fusion at $s^*$ is obtained as
\beqn
\tilde J(s^*)&=&\tilde J(s_0)\left(\frac{1-s^*}{1-s_0}\right)^{-n_b}=\tilde J(s_0)(r^*)^{-\frac{n_b}{n_1+n_b}} \nonumber \\
&=&\tilde J(s_0)\left(\frac{2n_b+n_1-1}{n_b+2n_1+1}\right)^{-\frac{n_b}{2(n_1+n_b)}}.
\eeqn
The energy gap at $s^*$ is then given by
\beqn
\epsilon=2\tilde J(s_0)\left(\frac{2n_b+n_1-1}{n_b+2n_1+1}\right)^{-\frac{n_b}{2(n_1+n_b)}} \nonumber \\
\times \left(3\frac{n_1+n_b}{n_b+2n_1+1}\right)^{1/2}.
\eeqn
From these, $\tau_c\epsilon$ is given by
\beqn \label{eq:tau_epsilon}
\tau_c\epsilon=\frac{1}{6}\left(\frac{2n_b+n_1-1}{n_b+2n_1+1}\right)^{\frac{n_1+n_b-1}{2(n_1+n_b)}} \nonumber \\
\times \frac{n_b+2n_1+1}{1-s_0}.
\eeqn

\section{Numerically exact calculations in one dimension}
\label{app:free-fermion}

We consider the RTFIM in one dimension with open boundaries and assume a general time-dependence of couplings and fields, which can be specified later. 
The eigenvalue problem of the instantaneous Hamiltonian
\be
{\cal H}(s) =
-\sum_{n=1}^{L-1}J_{n}(s)\sigma_n^x \sigma_{n+1}^x-\sum_{n=1}^Lh_n(s)\sigma_n^z,
\label{H}
\ee
where $s=t/\tau$, can be solved in a standard way \cite{pfeuty,lsm}.
First, by a Jordan-Wigner transformation
\beqn
c_n^{\dagger}+c_n&=&\left(\prod_{m<n}-\sigma_m^z\right)\sigma_n^x, \nonumber \\ 
c_n^{\dagger}-c_n&=&i\left(\prod_{m<n}-\sigma_m^z\right)\sigma_n^y, \qquad n=1,2,\dots,L \nonumber
\label{JW}
\eeqn
the Hamiltonian can be rewritten in a quadratic form in fermionic creation and annihilation operators, $c_n^{\dagger}$ and $c_n$, respectively: 
\beqn
{\cal H}&=&-\sum_{n=1}^{L-1}J_n(s)\left(c_n^{\dagger}-c_n\right)\left(c_{n+1}^{\dagger}+c_{n+1}\right) \nonumber \\
&&-\sum_{n=1}^Lh_n(s)\left(2c_n^{\dagger}c_n-1\right).
\label{quad}
\eeqn
Introducing Clifford operators as $a_n=c_n^{\dagger}+c_n$ and $b_n=c_n^{\dagger}-c_n$, the Hamiltonian in Eq.~(\ref{quad}) can be recast using a matrix notation as 
\be 
{\cal H}   =-\frac{1}{2}a^TMb, 
\label{Hmatrix}
\ee
where $a^T$ and $b$ denote a row vector of operators $a_n$ and a column vector of operators $b_n$, respectively. The bidiagonal matrix $M$ is given by 
\be
M=
-2\begin{bmatrix}
h_1(s) & & & & \cr
J_1(s) & h_2(s) & & & \cr
 & J_2(s) & \ddots & & \cr
 & & \ddots & h_{L-1}(s) & \cr
 & & & J_{L-1}(s) & h_L(s)
\end{bmatrix},
\label{M}
\ee
Then, a subsequent Bogoliubov-Valatin transformation
\be
\eta_k=\frac{1}{2}\sum_{n=1}^L\left[\left(\Phi_{nk}+\Psi_{nk}\right)c_n + \left(\Phi_{nk}-\Psi_{nk}\right)c_n^{\dagger}\right],
\label{Hc}
\ee
for $k=1,\dots,L$, with suitable real and orthogonal coefficient matrices $\Phi$ and $\Psi$ brings the Hamiltonian to a diagonal form
\be
{\cal H}=\sum_{k=1}^{L}\epsilon_k\left(\eta_k^{\dagger}\eta_k-\frac{1}{2}\right).
\label{diag}
\ee   
As can be seen from Eq.~(\ref{Hmatrix}), 
the coefficient matrices $\Phi$ and $\Psi$, as well as the excitation energies $\epsilon_i$ can be found by performing the singular-value decomposition of 
$
M=\Phi\mathcal{D}\Psi^T,
$
 where $\mathcal{D}={\rm diag}\{\epsilon_1,\epsilon_2,\dots,\epsilon_L\}$.

Next, we proceed to calculate the ratio in the adiabatic theorem, which can be written in the fermion language as 
\be 
Q(s)=\frac{\left|\langle 0|\frac{d{\cal H}(s)}{ds}\eta^{\dagger}_2\eta^{\dagger}_1|0\rangle\right|}{(\epsilon_1+\epsilon_2)^2}.
\label{atff}
\ee
Here, $|0\rangle$ denotes the vacuum of $\eta$ fermions, while $\eta^{\dagger}_2\eta^{\dagger}_1|0\rangle$ is the lowest-energy excited state in the even sector, with excitation energy $\epsilon_1+\epsilon_2$.
We can observe that 
$\frac{d{\cal H}(s)}{ds}$ is again an RTFIM Hamiltonian with couplings 
$\frac{dJ_n(s)}{ds}$ and fields $\frac{dh_n(s)}{ds}$ and as such, it can be written as $\frac{d{\cal H}}{ds}   =-\frac{1}{2}a^TM'b$ with 
$M'\equiv\frac{dM}{ds}$
This can be further expressed by $\eta$ fermions as 
$\frac{d{\cal H}}{ds}  =-\frac{1}{2}\alpha^T\Phi^TM'\Psi\beta$, where $\alpha^T$  is a row vector of Clifford operators 
$\eta^{\dagger}_n+\eta_n$, whereas $\beta$ is a column vector of Clifford operators $\eta^{\dagger}_n-\eta_n$. 
Using this, the matrix element can be evaluated to give 
\be 
\frac{1}{2}\left|[\Phi^TM'\Psi]_{12}-[\Phi^TM'\Psi]_{21}\right|.
\ee

\end{document}